\documentclass[12pt]{article}
\usepackage{epsf}

\catcode`\@=11

\def\section{\@startsection{section}{1}{\z@}{3.5ex plus 1ex minus
 .2ex}{2.3ex plus .2ex}{\bf}}

\def\thesubsection{\arabic{section}.\arabic{subsection}}
\renewcommand{\subsection}[1]{\addtocounter{subsection}{1}
\vspace{2.5mm}\par\noindent {\it \thesubsection . #1}\par
 \vspace{0.5mm} }
\catcode`\@=12
\newfont{\mbm}{msbm10 scaled\magstep1}
\def\bb#1{\hbox{\mbm #1}}

\def\reflist{\section*{References\markboth
        {REFLIST}{REFLIST}}\list
        {[\arabic{enumi}]\hfill}{\settowidth\labelwidth{[999]}
        \leftmargin\labelwidth
        \advance\leftmargin\labelsep\usecounter{enumi}}}
 \relax

\def\e{\epsilon}

\newcommand{\be}{\begin{equation}}
\newcommand{\ee}{\end{equation}}
\newcommand{\ba}{\begin{eqnarray}}
\newcommand{\ea}{\end{eqnarray}}

\catcode`\@=11

\begin{document}
\begin{titlepage}
\rightline{{CERN-TH/2002-091}}
\rightline{{CPHT RR 003.0202}}
\rightline{{LPT-ORSAY 02-37}}
\rightline{{LPTM 02-55}}
\rightline{{hep-th/0205096}}
\vskip 2cm
\centerline{{\large\bf Orientifolds of String Theory Melvin backgrounds}}
\vskip 1cm
\centerline{Carlo Angelantonj${}^\dagger$,
Emilian Dudas${}^{\ddagger,\star}$ and Jihad Mourad${}^{\ddagger,*}$}
\vskip 0.5cm
\centerline{\it ${}^\dagger$ TH-Division, CERN, CH-1211 Geneva 23}
\vskip 0.3cm 
\centerline{\it ${}^\ddagger$ Laboratoire de Physique Th{\'e}orique
\footnote{Unit{\'e} Mixte de Recherche du CNRS (UMR 8627).}}
\centerline{\it Univ. de Paris-Sud, B{\^a}t. 210, F-91405 Orsay Cedex}
\vskip 0.3cm 
\centerline{\it ${}^\star$ Centre de Physique Th{\'e}orique,
Ecole Polytechnique, F-91128 Palaiseau}
\vskip 0.3cm 
\centerline{\it ${}^*$ Laboratoire de Physique Th{\'e}orique et 
Mod{\'e}lisation,}
\centerline{\it
Univ. de Cergy-Pontoise, Site de Neuville III, F-95031
Cergy-Pontoise}

\vskip  1.0cm
\begin{abstract}

We study the dynamics of type I strings on Melvin backgrounds,
with a single or multiple twisted two-planes. We construct two
inequivalent types of orientifold models that correspond to
(non-compact) irrational versions of Scherk-Schwarz type
breaking of supersymmetry. In the first class of vacua, D-branes and O-planes
are no longer localized in space-time but are smeared along
the compact Melvin coordinate with a characteristic profile. 
On the other hand, the second class of orientifolds involves
O-planes and D-branes that are both rotated by an angle proportional
to the twist. In case of ``multiple Melvin spaces'', 
some amount of supersymmetry is recovered if the planes are twisted
appropriately and part of the original O-planes are transmuted into
new ones. The corresponding boundary and crosscap states are determined.
\end{abstract}
\end{titlepage}

\section{Introduction and summary of results}

The Melvin space \cite{Melvin}, \cite{Gibbons:1987wg,dowker} is a flat
$\bb{R}^9 \times S^1$ manifold 
subject to non-trivial identifications: whenever the compact coordinate
$y$ wraps $n$ times around the circle of length $2 \pi R$, 
the angular variable $\phi_0$ of a non-compact two-plane is rotated by
$2\pi n \gamma$ where $\gamma$ is the twist parameter\footnote{In
\cite{dm} $\gamma$ was denoted by $BR$, 
where $B$ is interpreted as a closed magnetic field.}. This is
summarized by
 
\be
(y,\phi_0)=(y+2\pi n R , \phi_0 + 2 \pi m + 2 \pi n \gamma  )
\ . \label{one}
\ee
{}From (\ref{one}) it is clear that in a bosonic theory twists differing by
integers define the same space-time and, as a result we can restrict
ourselves to the interval $\gamma \in [0, 1]$. Whenever fermions are
present, however, different backgrounds correspond to values of 
$\gamma \in [0,2]$. Actually, we can always introduce a true angular
variable $\phi=\phi_0 -{\gamma \over R} y$, but now the Melvin metric
reads
\be
ds^2=d\rho^2+\rho^2 \left(d\phi+{\gamma \over R}  dy\right)^2
+dy^2+d{\bf x}^2 \ , \label{c2}
\ee
where ${\bf x}$ denotes the seven spectator coordinates.
The geometric interpretation of the Melvin space is thus quite
simple. Under a $2 \pi$ rotation of the $y$ coordinate a spin-$j$
field $\Phi$ is subject to the monodromy
\be
\Phi (y+ 2 \pi R) =  e^{2 \pi i \gamma j} \; \Phi (y) \, . \label{i1}
\ee
As a result, the Melvin model (\ref{one}) can be well interpreted as
a non-compact version of the Scherk-Schwarz mechanism \cite{ss,ssclosed}, 
with the only (important) difference that the parameter $\gamma$
can now be any irrational number.

 A full-fledged description of String Theory on this Melvin
background is actually possible \cite{Russo:1995tj}. 
Recently, there has been an  increasing interest in
various aspects of these Melvin backgrounds in Kaluza-Klein reductions
\cite{Gibbons:1987wg,dowker}, in M-theory, in Type II and in heterotic 
strings \cite{Costa:2001nw}-\cite{Takayanagi:2001jj}.
D-brane probes in such backgrounds have been analysed in
\cite{dm} and \cite{tu2}, while Melvin spaces also offer an interesting
arena to discuss the fate of closed-string tachyons \cite{eg}-\cite{dghm}. 

The purpose of this paper is to construct (non-supersymmetric and
supersymmetric) orientifolds \cite{cargese} (see, for reviews, 
\cite{ad,review}) of the type IIB Melvin backgrounds. We
construct inequivalent orientifold models, 
study their spectra and determine their boundary and crosscap states. 
We find that the models  share some interesting
properties of the orientifolds of the Scherk-Schwarz models
constructed in \cite{ads1,ssopen}. We give a 
geometrical interpretation of the orientifold planes and D-branes present in
the models and find some new interesting phenomena. 
In particular, in a first class of orientifolds, O-planes and D-branes 
are no longer localized in $y$. Rather, they are smeared along the $S^1$
with a characteristic profile.
The second class of orientifolds are also peculiar, and
describe rotated O-planes and D-branes, as well as the generation
of new O-planes, phenomena that, to the best of our knowledge, did not 
emerge before.

This paper is organized as follows. In section 2 we review
the quantization of closed superstrings in the 
Melvin background using NSR fermions. In section 3 we construct
the Klein bottle, annulus and M{\"o}bius amplitudes pertaining
to the first class of orientifolds. Then, in
section 4, we give a geometrical interpretation of the branes and O-planes:
they are described by a non-trivial wave function which is supported on
a discrete set of points in $S^1$. We also derive the coupling of 
closed-string states to the crosscap and D-branes.
Section 5 deals with the second class of orientifold models, that involve
rotated pairs of orientifold plane-antiplanes. In this case, the computations
of massless tadpoles is quite subtle and we resort to boundary-states
techniques and quantum mechanics analysis to extract them. 
Section 6 briefly discusses the issue of closed- and open-string 
tachyons, and the fate of the orientifold vacua. Finally, sections 7,8 and 9
deal with double Melvin backgrounds and with their orientifolds. In this
case, whenever the twist parameters are chosen to be the same, (half of)
the original supersymmetry
is recovered in the string spectrum, and new orientifold planes are 
generated. This phenomenon is reminiscent of the brane transmutation of
\cite{mag}. The appendix defines and collects the crosscap and boundary 
states used in the paper.

\section{Closed strings in Melvin backgrounds}

In this section we review some known facts about closed string dynamics
in Melvin backgrounds. It will serve the purpose of introducing the 
notation and derive their partition functions that, in the spirit of
\cite{cargese} are the starting point for the orientifold construction.
Unless explicitly stated, in the following we shall assume an irrational
twist $\gamma$, though special cases will be discussed as well. 

According to eq. (\ref{c2}), the string dynamics in the Melvin space-time 
is governed by the world-sheet Action
\be
S=- {{1}\over{4\pi\alpha'}}\int \left[ d\rho\wedge ^{*}d\rho+
\rho^2 \left(d\phi+{\gamma \over R}  dy\right)
\wedge ^{*} \left( d\phi+{\gamma \over R}  dy \right)
+dy\wedge ^*dy \right] \, . \label{c3}
\ee
This Action has a very simple equivalent description after we introduce
the complex coordinate
\be
Z_0=\rho \, e^{i(\phi+{\gamma \over R}  y)}   \,, \label{c4}
\ee
that actually corresponds to a free boson
\be
\left( {\partial^2 \over \partial\sigma^2} - {\partial^2 \over \partial
\tau^2} \right) Z_0 =0 \,,
\ee
with twisted boundary conditions
\be
Z_0(\sigma+2\pi) = e^{2\pi i n\gamma} \ Z_0(\sigma) \, , \label{c7}
\ee
where $n$ is the winding mode in the $y$ direction.
The compact $y$-coordinate also corresponds to a free field, and, as usual,
satisfies the periodicity condition
\be
y(\sigma + 2 \pi) = y(\sigma ) + 2 \pi n R
\ee
for a circle of length $2 \pi R$, while its conjugate momentum 
\be
P_y={1 \over 2\pi \alpha'}\int d\sigma\partial_\tau y + 
{\gamma \over R} L\ =\ {k \over R} \, ,
\label{ccc1}
\ee
quantized as usual for Kaluza-Klein (KK) modes, now
receives contributions from the angular momentum $L$ on the $(\rho,\phi)$ 
plane. 

As a result the mode expansion of the $y$ coordinate is affected in its
zero-mode contributions, and reads
\ba
y &=& y_0 + nR\sigma + \alpha' \, {k -\gamma L\over R} \, \tau
\nonumber \\
& & + \sqrt{\alpha'} \sum_{m=1}^{\infty}
{1\over \sqrt{m}} \left[ {y_m }e^{-im\sigma_+}+
{y_m^\dagger}e^{im\sigma_+} 
+{\tilde y_m }e^{-im\sigma_-}+
{\tilde y_m^\dagger }e^{im\sigma_-}\right] \, , \label{c8}
\ea
where we have introduced the light-cone coordinates 
$\sigma_\pm = \tau \pm \sigma$, while the normalization of the oscillators 
has been chosen in order to have 
conventional (in quantum mechanics) canonical commutation relations 
$[y_m,y_n^{\dagger}]=\delta_{mn}$, $[{\tilde y}_m, {\tilde y}_n^{\dagger}]
=\delta_{mn}$. 

Given the boundary condition of eq. (\ref{c7}),
the Fourier modes of the $Z_0$ coordinate have shifted frequencies, 
as pertains to twisted fields
\footnote{Notice that we changed notations
with respect to \cite{dm} ($\tilde a_m \leftrightarrow \tilde
b_m$).},
\ba
Z_0^{(\nu)} (\sigma,\tau) &=& \sqrt{\alpha'} \left[
\sum_{m=1}^{\infty}{{a_m \over
\sqrt{m-\nu}}}e^{-i(m-\nu)\sigma_+}+\sum_{m=0}^{\infty}
{b_m^{\dagger} \over{\sqrt{m+\nu}}}e^{i(m
+\nu)\sigma_+} \right. 
\nonumber\\
& & \left. +  \sum_{m=0}^{\infty}{{\tilde b_m \over
\sqrt{m+\nu}}}e^{-i(m+\nu)\sigma_-}+\sum_{m=1}^{\infty}
{\tilde a_m^{\dagger} \over{\sqrt{m-\nu}}}e^{i(m
-\nu)\sigma_-}\right] \, . \label{expa}
\ea
As for the $y$ coordinate, we have normalized the Fourier modes so that 
the string oscillators satisfy canonical commutation relations.
Moreover, we have introduced the variable $\nu=f(\gamma n)$, where
the function $f$ is  1-periodic and $f(x)=x$ for $0<x<1$. 
More explicitly, $\nu=n\gamma -[n\gamma ]$ for $n$ 
positive and  $\nu=n\gamma -[n\gamma ]+1$ for $n$ negative.
In (\ref{expa}), $a_n^\dagger$ and $b_n^\dagger$ are creation
operators.

We can now use the previous expansions to derive the left- and right-handed
components of the angular momentum $L\equiv J+\tilde J$ of the $(\rho , \phi)$
plane,
\be
J = - b_0^{\dagger} b_0
+\sum_{m=1}^{\infty}( a_m^\dagger a_m - b_m^\dagger b_m) \, , \qquad
{\tilde J} = \tilde b^{\dagger}_0\tilde b_0
-\sum_{m=1}^{\infty}( \tilde a_m^\dagger\tilde a_m
- \tilde b_m^\dagger\tilde b_m) \, , \label{angm}
\ee
as well as the total Hamiltonian 
\ba
H &=& L_0 + {\bar L}_0 
\nonumber\\
&=& N+\tilde N-\nu(J-\tilde J)+
{\alpha' \over 2 }\left[ \left( {nR \over\alpha'} \right)^2
+\left({ k -\gamma (J+\tilde J)  \over R}\right)^2 \right]
- {\textstyle{1 \over 4}}(1-2\nu)^2 \, , \label{mass}
\ea
pertaining to the compact ($y$) and twisted ($Z_0$) coordinates.
Here, $N$ ($\tilde N$) denotes the total number operator of left (right) 
oscillators. 

For vanishing $\nu$, $Z_0$ describes a conventional complex coordinate
with periodic boundary conditions. As a result, it acquires 
zero-mode contributions, while the associated Hamiltonian reduces to
the familiar one corresponding to a pair of free bosonic fields.

We now have all the ingredients necessary to compute the partition function 
pertaining to bosonic strings moving on a Melvin background
and, for simplicity we shall neglect the contributions 
from the remaining 21 free coordinates. 
After a Poisson resummation on $k$, needed to linearize the $J$ dependence
of the left and right Hamiltonians, and omitting the explicit 
$\tau$-integration, $\int d^2 \tau /\tau_2$, the torus amplitude reads
\be
{\cal T} = {R\over \sqrt{\alpha'\tau_2}}{1 \over |\eta|^2}
\sum_{\tilde k, n}e^{-{\pi R^2 \over \alpha'\tau_2}
|\tilde k+\tau n|^2} {\rm tr}_\nu \left[ 
e^{-2\pi i\gamma  \tilde k(J+\tilde J)}
q^{L_0}\bar q^{{\bar L}_0}\right] \ ,
\ee
where ${\rm tr}_{\nu}$ is the trace in the $\nu$-twisted sector.

As a result, standard (orbifold-like) calculations lead to the final result
\be
{\cal T}={R\over \sqrt{\alpha'\tau_2}}{1 \over |\eta|^2}
\sum_{\tilde k, n}{\cal T} (\tilde k,n)\, 
e^{-{\pi R^2 \over \alpha'\tau_2}
|\tilde k+\tau n|^2}   \, , \label{tor}
\ee
with
\be
{\cal T} (0,0) = {v_2 \over \tau_2}{1 \over |\eta|^4} \qquad {\rm and}
\qquad
{\cal T} (\tilde k,n) =
\left| {\eta \over \vartheta \left[{1/2+\gamma  n \atop 1/2+\gamma  \tilde k} 
\right]} \right|^2 \, . \label{part}
\ee

Here and in the following we shall always introduce the 
dimensionless volume 
\be
v_d={V_d \over (4\pi ^2 \alpha')^{d/2}} \,.
\ee

Before turning to the superstring case, let us pause for a moment and remind
that for a non-vanishing twist the zero-modes contribute to ${\cal T}$ with
a factor 
\be
\int d^2 p \; \langle p | e^{-2\pi i L \gamma \tilde k} |p\rangle \, 
(q\bar q)^{\alpha ' p^2 /4} = {1\over {\rm det} (1- e^{- 2 \pi i
\gamma {\tilde k}})}
= {1\over [2 \sin (\pi \gamma \tilde k )]^2} \,,
\label{zerom1}
\ee
that in standard orbifold compactifications is responsible for the fixed-point 
multiplicities in the twisted sectors. Although this expression is clearly
well-defined for any twist $0< \gamma <1$, it can actually be extended to
comprise the two extrema if the limit $\gamma \to 0,1$ is taken
before the integral is effectively evaluated. In this case, the result
is proportional to $v_2 / \tau_2$, as pertains to the zero-modes of an
un-twisted, non-compact, free boson. 

The inclusion of the twisted world-sheet fermions $\lambda$, partners
of $Z_0$, is then straightforward. Their mode expansion is
\ba
\lambda_{+}(\sigma_+) &=&
\sum_{m=1}^{\infty} \mu_m e^{-i(m-\nu')\sigma_+}
+\sum_{m=0}^{\infty} \zeta_m^\dagger e^{i(m+\nu')\sigma_+} \, , \nonumber
\\
\lambda_{-}(\sigma_-) &=&
\sum_{m=1}^{\infty} \tilde \mu_m^\dagger e^{i(m-\nu')\sigma_-}
+\sum_{m=0}^{\infty} \tilde\zeta_m e^{-i(m+\nu')\sigma_-} \, ,
\ea
where $\nu'=\nu$ in the R sector and $\nu'= f( {1 \over 2} +\nu)$ in the
NS one.

Their Hamiltonians and angular momenta are then given by
\ba
L_f&=&\sum_{m=1}^{\infty}(m-\nu')\mu_m^\dagger \mu_m + 
\sum_{m=0}^{\infty}(m+\nu')\zeta^\dagger_m\zeta_m +a_f \, ,
\nonumber\\ 
\bar L_f&=& \sum_{m=1}^{\infty}(m-\nu')\tilde \mu_m^\dagger\tilde  \mu_m  
+ \sum_{m=0}^{\infty}(m+\nu') {\tilde \zeta}^\dagger_m {\tilde
  \zeta}_m + {\tilde a}_f \, , 
\ea
and 
\ba
J_f&=&-\zeta_0^\dagger\zeta_0+
\sum_{m=1}^{\infty} 
(\mu_m^\dagger \mu_m-\zeta^\dagger_m\zeta_m)+b \ , \nonumber \\
{\tilde J}_f&=&\tilde\zeta_0^\dagger\tilde\zeta_0-
\sum_{m=1}^{\infty} (\tilde \mu_m^\dagger \tilde\mu_m-\tilde
\zeta^\dagger_m\tilde\zeta_m) + \bar b \, . \label{angf}
\ea
Notice that the normal ordering constants $b$ and $\bar b$  
(with $b=-\bar b= {1 \over 2} -\nu'$) are fixed by modular invariance of 
the corresponding partition function (\ref{parti}). 

Combining the contributions of the bosonic and fermionic coordinates then
yields the total right- and left-handed Hamiltonians
\ba
L_0 &=& L_f + N_b + {\alpha' \over 4}
\left( {k \over R}-{\gamma \over R} 
(J+\tilde J) + {nR\over \alpha'} \right)^2
-\nu J_b +a_b \, , 
\nonumber \\
\bar L_0 &=& \bar L_f+ {\bar N}_b + {\alpha' \over 4}
\left( {k \over R}-{\gamma \over R} (J+\tilde J) - {nR\over \alpha'}
\right)^2 +\nu \tilde J_b+\tilde a_b \, , \label{c9}
\ea
where the overall normal ordering constants are 
$a({\rm NS} )={1\over 2} (\nu-1)$ when $\nu< {1\over 2}$,
$a({\rm NS})=- {1\over 2} \nu$ when $\nu>{1\over 2}$ and 
$a({\rm R})=0$. 

Resorting to standard orbifold-like calculations, and after a Poisson 
resummation on the KK momenta $k$, the torus amplitude for the type 
IIB superstring
in the Melvin background reads
\be
{\cal T} = {R \over \sqrt{\alpha ' \tau_2}} \, 
{v_7 \over \tau_2^{7/2} \, |\eta|^{12}}
\sum_{\tilde k, n}{\cal T} (\tilde k,n) \,
e^{-{\pi R^2 \over \alpha'\tau_2}
|\tilde k+\tau n|^2}   \, , \label{tor2}
\ee
with
\ba
{\cal T} (0,0) &=& {v_2 \over \tau_2\, |\eta|^4}
\left| \sum_{\alpha,\beta} {\textstyle {1\over 2}}\, \eta_{\alpha\beta}
{\vartheta^4 \left[{\alpha \atop \beta} \right] \over \eta^4}
\right|^2 \,,
\label{parti} \\
{\cal T} (\tilde k,n) &=& \left| \sum_{\alpha,\beta} 
{\textstyle{1\over 2}}\, \eta_{\alpha
\beta} e^{-2\pi i \beta \gamma n} {\vartheta^3 \left[{\alpha \atop \beta}
\right] \over \eta^3}
{\vartheta \left[{\alpha+\gamma n \atop \beta+\gamma \tilde k} \right]
\over \vartheta \left[ {1/2+\gamma n \atop 1/2+\gamma \tilde k} \right]}
\right|^2 \,. \nonumber
\ea
Here, $\eta_{\alpha\beta}= \bar \eta _{\alpha \beta} = 
(-1)^{2\alpha+2\beta + 4 \alpha \beta}$
defines the GSO projection in the IIB case\footnote{The analogous 
amplitude for the IIA superstring
is easily obtained introducing the proper GSO projection $\eta_{\alpha \beta}
= (-1)^{2\alpha + 2 \beta + 4 \alpha \beta}$ and $\eta_{{1\over 2}\, 
{1\over 2}} = - \bar \eta_{{1\over 2}\, {1\over 2}}$.}.

Notice that there are scalar states compatible with the GSO 
projection belonging to the NS-NS sector, with  
zero KK momentum and one unit of windings, $n=\pm 1$. Their mass
is given by 
\be 
M^2 = {2\over \alpha '} (L_0 + \bar L_0 ) = \left({R\over \alpha'}\right)^2 -
{2 \gamma \over \alpha'} \label{tmass}
\ee
and, as a result, a tachyon is present in the spectrum for
$R^2 < 2\alpha' \gamma $. 

As usual, string theory affords new interesting vacuum configurations
whenever momenta and windings are interchanged. Generically, this corresponds 
to T-dualities, that for the Melvin background we are dealing with 
are {\it not} a symmetry. Hence, a Buscher duality on $y$ 
\be
dy \to dy' = \left( 1 + {\gamma^2 \rho^2 \over R^2}\right) \, ^* dy +
{\gamma \rho^2 \over R} \, ^* d \phi \,, \label{tildey}
\ee 
yields the new background
\be
S=- {{1}\over{4\pi\alpha'}}\int \left[ 
d\rho\wedge ^{*}d\rho+
{R^2 \over R^2+ \gamma^2 \rho^2} \left( \rho^2 d\phi \wedge ^{*} d\phi
+ d{ y'}\wedge ^*d{ y'} - 
2 {\gamma \rho^2 \over R}  d \phi \wedge d { y'} \right) \right] 
\,. \label{buscher}
\ee
The space-time of eq. (\ref{buscher}) is now curved, and involves 
a non-vanishing $B$-field, and a varying dilaton
\be
e^{\Phi_0}=  {g_s \over {\sqrt{1+\gamma^2\rho^2/R^2}}} \,.
\ee 

The quantization of the $\sigma$-model (\ref{buscher}) leads then
to the IIB partition function
\be
{\cal T} = { \sqrt{\alpha'} \over R \sqrt{\tau_2}} \,
{v_7 \over \tau_2^{7/2} \, |\eta|^{12}}
\sum_{k,\tilde n} {\cal T} (k, \tilde n ) \,
e^{-{\pi \alpha' \over R^2 \tau_2}
|\tilde n + \tau k|^2}  \, , \label{tor3}
\ee
with
\ba
{\cal T} (0,0) &=& {v_2 \over \tau_2\, | \eta|^4}
\left| \sum_{\alpha,\beta} {\textstyle{1\over 2}} \, \eta_{\alpha\beta}
{\vartheta^4 \left[{\alpha \atop \beta} \right] \over \eta^4}
\right|^2  \,,
\label{parti2} \\
{\cal T} (k, \tilde n ) &=& \left|
\sum_{\alpha,\beta} {\textstyle{1\over 2}} \, 
\eta_{\alpha \beta}e^{-2\pi i \beta \gamma  k }
{\vartheta^3 \left[ {\alpha \atop \beta}\right] \over \eta^3}
{\vartheta \left[ {\alpha+\gamma k \atop \beta + \gamma \tilde n} \right]
\over \vartheta \left[ {1/2+\gamma k \atop 1/2+\gamma \tilde n} \right]}
\right|^2 \,.
\nonumber
\ea

Although both (\ref{tor2}) and (\ref{tor3}) models are formally defined
for $0<\gamma <1$, they can be extended to include
the limiting cases $\gamma \to 0,1$
if zero-mode contributions are handled with care. In this cases, both 
amplitudes (\ref{tor2}) and (\ref{tor3}) reproduce the type IIB superstring
for $\gamma =0$, while they yield the Scherk-Schwarz and M-theory breakings
of \cite{ads1} for $\gamma =1$. As we shall see in the following
sections, the same properties are shared by their orientifolds.

\section{Orientifolds of Melvin backgrounds}

The background we are considering has the nice property of 
preserving the invariance of the IIB string under the world-sheet
parity $\Omega$. We can then proceed to construct the corresponding
orientifolds, that, as we shall see, have in store interesting 
results. 

The action of the world-sheet parity on the relevant coordinates $y$ 
and $Z_0$ of this Melvin background can be easily retrieved from similar 
toroidal and orbifold constructions
\be
\Omega \; y(\sigma) \; \Omega^{-1} = y (-\sigma) \, , \qquad 
\Omega \; Z^{(\nu)}_0 (\sigma) \; \Omega^{-1} = Z^{(1-\nu)}_0 (-\sigma) \,,
\label{o01}
\ee
and implies
\be
\Omega N \Omega^{-1} = \tilde N+(\tilde J-\tilde b)
 \, , \qquad \Omega \tilde N
\Omega^{-1} = N-(J-b) \, ,
\qquad \Omega J \Omega^{-1}= {\tilde J} \, . \label{o2}
\ee
while, as usual, only zero-winding states survive the projection.
The operation (\ref{o2}) implies  
$\Omega a_{m}\Omega^{-1}= 
\tilde b_{m-1} \ , \ \Omega b_{m} \Omega^{-1} = {\tilde a}_{m+1}$ 
on the $Z_0$ oscillators and
therefore it maps string oscillators into Landau
levels and vice versa.  Moreover,  the states $|K\rangle$ 
propagating in the
Klein bottle must satisfy the condition $(J - {\tilde J}) |K\rangle = 0 $. 

The Klein-bottle amplitude is then given by
\be
{\cal K}= {\textstyle{1 \over 2}} {\rm tr} \; \left( \Omega \ 
q^{L_0}\bar q^{\bar L_0} \right) = 
{\textstyle {1 \over 2 }} 
 {\rm tr} \;  \left( (q^2) ^{ (H + {\alpha' \over 4} Q^2)} \right)
\ , \label{o3}
\ee
where $q=e^{- 2\pi \tau_2}$ and, as usual, the amplitude depends on 
the modulus of the doubly-covering torus $2 i \tau_2$. Moreover, $Q$ denotes 
the non-compact momenta, and
\be
H = N + {\alpha' \over 4} \left( {k \over R} - {\gamma \over
R} (J+{\tilde
J}) \right)^2 + a \, , \label{o4}
\ee
is the $\Omega$-invariant Hamiltonian, with $a$ vanishing in the R-R sector
and equal to $-{1\over 2}$ in the NS-NS one.
Here and in the following we have omitted the modular integral
$\int_0^\infty d \tau_2 / \tau_2$.

To explicitly compute the trace in (\ref{o4}) it is more convenient to 
perform a Poisson resummation over the KK momenta to linearize
the $J$ dependence in $H$. As a result,
\be
{\cal K} = {\textstyle {1\over 2}} {R \over\sqrt{\alpha' \tau_2}} 
\sum_{\tilde k}\; {\rm tr} \; \left( e^{- 2 \pi i {\tilde k} 
\gamma (J+{\tilde J}) -  
4 \pi \tau_2  (N+a + {\alpha ' \over 4} Q^2)} 
e^{-{\pi \over \alpha' \tau_2}  ( R{\tilde k})^2 } \right) \,, \label{o5}
\ee
and, explicitly,
\be
{\cal K} = {\textstyle{1\over 2}} {R \over \sqrt{\alpha ' \tau_2}} \, 
{ v_7 \over \tau_2^{7/2} \eta^6} \sum_{\tilde k} e^{- {\pi \over \alpha ' 
\tau_2} (R \tilde k)^2}\, {\cal K} (\tilde k )\,, \label{kdirm}
\ee
where
\ba
{\cal K} (0) &=& {v_2 \over \tau_2 \eta^2} \sum_{\alpha ,\beta} 
{\textstyle{1\over 2}} \, \eta_{\alpha \beta} {\vartheta^4 \left[ {\alpha 
\atop \beta} \right] \over \eta^4} \,,
\nonumber \\
{\cal K} (\tilde k) &=& - {2 \sin (2 \pi \gamma \tilde k ) \over 
\left[ 2 \sin (\pi \gamma \tilde k )\right]^2} \sum_{\alpha ,\beta} 
{\textstyle{1\over 2}} \, \eta_{\alpha \beta} {\vartheta^3 \left[ {\alpha 
\atop \beta} \right] \over \eta^3} {\vartheta \left[ {\alpha \atop
\beta + 2 \gamma \tilde k} \right] \over \vartheta \left[ {1/2 \atop
1/2 + 2 \gamma \tilde k} \right] } \,. \label{kdirmb}
\ea
Notice the peculiar behaviour of the string oscillators in ${\cal K}$.
They actually feel a ``double rotation'' as can be easily deduced from
the explicit expression of the Hamiltonian (\ref{o4}), that is a 
straightforward consequence of the vertical doubling of the elementary 
cell representing the Klein-bottle surface.

Before turning to the transverse channel some comments are in order. 
While for generic values of the twist $\gamma$ a proper particle 
interpretation is not transparent in ${\cal K}$, as due to the non-standard
dependence of the lattice contribution on $\tau_2$, particularly 
interesting are the cases $\gamma = 0,1$. Although for these values 
${\cal K} (\tilde k )$ is na{\"\i}vely divergent, a careful analysis of the
zero-modes (\ref{zerom1}) reveals that the divergent term
is actually proportional to the infinite volume of the $Z_0$ two-plane:
\be
\lim_{\gamma \to 0,1} (\hbox{zero-modes}) = \int d^2 p \, 
\lim_{\gamma \to 0,1} \langle p | e^{-2\pi i L \gamma \tilde k} |p\rangle \, 
(q\bar q)^{\alpha ' p^2 /4} = {v_2 \over \tau_2} \,,
\ee
so that
\be
{\cal K} = {\textstyle{1\over 2}} \, {v_7 \over \tau_2^{7/2} \eta^6} \; 
{\cal K} (0) \; \sum_k \, q^{{\alpha ' \over 2} \left({k \over R}\right)^2} \,,
\ee
{\it i.e.} it reproduces the amplitude pertaining to the 
standard circle reduction of the
type I superstring ($\gamma= 0$) and to its nine-dimensional Scherk-Schwarz
compactification ($\gamma = 1$) \cite{ads1}.

We can now turn to the transverse-channel describing the closed-string 
interactions with the orientifold planes. An $S$ modular transformation
exchanges the two characteristics of the theta-functions while the 
lattice contribution has automatically the correct dependence on the
transverse-channel proper time $ \ell = 1/2\tau_2$. The corresponding
amplitude then reads:
\be
\tilde {\cal K} = {2^4\over 2} \; {R \over \sqrt{\alpha '}} \;
{v_7 \over \eta^6} \; \ell \; \sum_n \tilde{\cal K} (n) \; 
q^{(nR)^2 \over \alpha '} \,, \label{o7}
\ee 
with
\ba
\tilde{\cal K} (0) &=& {2\, v_2 \over \eta^2} \; \sum_{\alpha,\beta} 
{\textstyle{1\over 2}} \, \eta_{\alpha \beta} {\vartheta^4 \left[ {\alpha \atop
\beta} \right] \over \eta^4} \,,
\nonumber \\
\tilde{\cal K} (n) &=& {2 \sin (2 \pi \gamma n ) \over \left[
2 \sin (\pi \gamma n ) \right]^2} \; \sum_{\alpha,\beta} {\textstyle{1\over 2}}
\, \eta_{\alpha \beta} e^{- 2 \pi i \gamma n (2 \beta -1)} 
{\vartheta^3 \left[ {\alpha \atop \beta} \right] \over
\eta^3} {\vartheta \left[ {\alpha + 2 \gamma n \atop \beta} \right] \over
i \ \vartheta \left[ {1/2 + 2 \gamma n \atop 1/2} \right]} \,.
\ea
The theta and eta functions depend on the modulus $i\ell$ of the
transverse-channel surface, and, as usual, the integration
$\int_0^\infty d\ell /\ell$ has been left implicit. 

One can then extract the contributions of $\tilde{\cal K}$ to 
NS-NS and R-R tadpoles. These can be associated to standard O9 
planes and thus require the introduction of 32 D9 branes, to which
we now turn.

The open-string Hamiltonian
\be
H = N + \alpha'  \left( {k \over R}- {\gamma \over R}  J\right)^2 +a \ , 
\label{o22}
\ee
determines the direct-channel annulus amplitude
\be
{\cal A} = {\textstyle{1\over 2}} \, {R \over \sqrt{\alpha ' \tau_2}}
\; {v_7 \over \tau_2^{7/2} \eta^6}\; \sum_{\tilde k} \;
N^2_{\tilde k} \, {\cal A} (\tilde k) \, e^{- {\pi \over \alpha' \tau_2}
(R \tilde k )^2} \,, \label{mand}
\ee
with
\ba
{\cal A} (0) &=& {v_2 \over \tau_2 \eta^2} \sum_{\alpha ,\beta} {\textstyle
{1\over 2}} \eta_{\alpha \beta} {\vartheta^4 \left[ {\alpha \atop \beta}
\right] \over \eta^4} \,,
\nonumber \\
{\cal A} (\tilde k) &=& - {1 \over 2 \sin (\pi \gamma \tilde k)}
\sum_{\alpha , \beta} {\textstyle{1\over 2}} \, \eta_{\alpha \beta}
{\vartheta^3 \left[ {\alpha \atop \beta}\right] \over \eta^3} \,
{\vartheta \left[ {\alpha \atop \beta + \gamma \tilde k}\right] \over
\vartheta \left[ {1/2 \atop 1/2 + \gamma \tilde k}\right]} \,,
\ea
depending on the imaginary modulus ${1\over 2} i \tau_2$ of the 
doubly-covering torus, and the M{\"o}bius amplitude
\be
{\cal M} = - {\textstyle{1\over 2}} \, {R \over \sqrt{\alpha ' \tau_2}}
\; {v_7 \over \tau_2^{7/2} \eta^6}\; \sum_{\tilde k} N_{2\tilde k} \,
{\cal M} (\tilde k) \, e^{- {\pi \over \alpha ' \tau_2} (R \tilde k )^2}
\,, \label{mmed}
\ee
with
\ba
{\cal M} (0) &=& {v_2 \over \tau_2 \eta^2} \sum_{\alpha , \beta} 
{\textstyle{1\over 2}} \, \eta_{\alpha \beta} {\vartheta^4 \left[
{\alpha \atop \beta}\right] \over \eta^4} \,,
\nonumber \\
{\cal M} (\tilde k) &=& - {1\over 2\sin (\pi \gamma \tilde k)} \,
\sum_{\alpha ,\beta} {\textstyle {1\over 2}} \, \eta_{\alpha \beta}
{\vartheta^3 \left[{\alpha \atop \beta}\right] \over \eta^3}
{\vartheta \left[ {\alpha \atop \beta + \gamma \tilde k}\right] \over
\vartheta \left[ {1/2 \atop 1/2 + \gamma \tilde k}\right] } \,,
\ea
depending on the complex modulus ${1\over 2} + {1\over 2} i \tau_2$ of its 
doubly-covering torus. Here $N_{\tilde k}$ counts the number of D9 branes
and we have considered the possibility of introducing particular Wilson 
lines or, in the T-dual language, of separating the branes.

The open-string spectrum is not transparent in eqs. (\ref{mand})
and (\ref{mmed}) due to the non-standard KK contributions. However, from 
the explicit expression of the Hamiltonian (\ref{o22}) it is evident that all 
fermions and scalars associated to components of the gauge vectors along
the $Z_0$ plane get a mass proportional to $\gamma^2 / R^2$, while no
tachyons appear in this open-string spectrum. 

$S$ and $P$ modular transformations and the redefinition 
$\ell = 2 /\tau_2$ in the annulus and $\ell = 1 /(2 \tau_2)$ in M{\"o}bius yield
the transverse-channel amplitudes
\be
\tilde {\cal A} = {2^{-4} \over 2} \, {R \over \sqrt{\alpha '}} \; {v_7 \over
\eta^6}\; \ell \; \sum_n N_n^2 \; \tilde{\cal A} (n) \;
q^{(nR)^2 \over 4 \alpha '} \,, \label{mannt}
\ee
with
\ba
\tilde{\cal A} (0) &=& {2^{-1} \, v_2 \over \eta^2} \sum_{\alpha ,\beta}
{\textstyle{1\over 2}} \, \eta_{\alpha \beta} {\vartheta^4 \left[ {\alpha
\atop \beta} \right] \over \eta^4} (i\ell) \,,
\nonumber \\
\tilde{\cal A} (n) &=& {1\over 2 \sin (\pi \gamma n )} \, \sum_{\alpha ,\beta}
{\textstyle{1\over 2}} \, \eta_{\alpha \beta} e^{- \pi i \gamma n (2
\beta -1)} \, {\vartheta^3 \left[
{\alpha \atop \beta}\right] \over \eta^3} \, {\vartheta \left[ {\alpha +
\gamma n \atop \beta} \right] \over i \ \vartheta \left[ {1/2 +
\gamma n \atop 1/2} \right] } (i\ell) \,,
\ea
and
\be
\tilde {\cal M} = - {R \over\sqrt{\alpha '}} \, {v_7 \over \eta^6} \;
\ell \; \sum_n N_{2n} \, \tilde{\cal M} (n) \, q^{(nR)^2 \over \alpha'}
\,,
\ee
with
\ba
\tilde{\cal M} (0) &=& {v_2 \over \eta^2} \, \sum_{\alpha , \beta} 
{\textstyle{1\over 2}} \, \eta_{\alpha \beta} {\vartheta^4 \left[
{\alpha \atop \beta} \right] \over \eta^4 } ({\textstyle{1\over 2}} +
i \ell) \,,
\nonumber \\
\tilde{\cal M} (n) &=& {1\over 2\sin (\pi \gamma n)} \, \sum_{\alpha , \beta}
{\textstyle{1\over 2}}\, \eta_{\alpha \beta} e^{- 2 \pi i \gamma n (2
\beta -1)} {\vartheta^3 \left[ {\alpha \atop \beta} \right] \over \eta^3}
{\vartheta \left[
{\alpha + 2\gamma n \atop \beta - \gamma n} \right] \over i \
\vartheta \left[
{1/2 + 2\gamma n \atop 1/2 - \gamma n} \right]} ({\textstyle{1\over 2}} +
i \ell) \,.
\ea
Notice the nice factorization of the zero-mode contribution to the 
transverse-channel M{\"o}bius amplitude
\be
{1 \over{\sin2\pi \gamma n}} \times {\sin {2 \pi \gamma n} \over \sin^2 {\pi
\gamma n}} = {1 \over{\sin^2 \pi \gamma n}} \,,  \label{o21}
\ee
that indeed is the geometric mean of those to $\tilde {\cal K}$
and $\tilde {\cal A}$.

Finally, tadpole cancellation determines the Chan-Paton multiplicities. 
In the simplest case of global cancellation one is led to the parametrization
$N_{n}= N= 32$ and to an SO(32) gauge group, while a local vanishing
of NS-NS and R-R tadpoles calls for the introduction of Wilson lines
\be
N_{2n} = N_1 + N_2 \,, \qquad N_{2n+1} = N_1 - N_2 \,, \label{o29}
\ee
with $N_1 = N_2 = 16$, thus yielding an ${\rm SO} (16) \times 
{\rm SO} (16)$ gauge group.

Also for these amplitudes one can consider the limiting cases $\gamma = 0,1$.
As expected, the former yields the standard nine-dimensional 
type I superstring with gauge group SO(32) (or ${\rm SO} (16) \times 
{\rm SO} (16)$ for local tadpole cancellation), while the $\gamma=1$ 
case reproduces the Scherk-Schwarz compactification of \cite{ads1}.

\section{Geometrical interpretation }

Let us pause for a moment and try to give a geometric description of the
orientifolds we just computed. As usual in this case, it is more convenient
to T-dualize the compact $y$ coordinate, whereby O9 planes and D9 branes 
are traded to pairs of O8 planes and corresponding D8 branes.

To this end, we resort  to boundary-state techniques, combined with
the informations encoded in the amplitudes (\ref{o7}) and (\ref{mannt}). 
In terms of the T-dual $y'$ coordinate of eq. (\ref{tildey}), 
the zero-mode contribution to boundary ($\alpha 
= b$) and crosscap ($\alpha =c$) states defined in the appendix
can then be written as
\be
|\alpha\rangle ={\cal N}^\alpha_0|k=0\rangle \otimes \ |P_Z=0\rangle +
\sum_{k\ \neq 0} {\cal N}_k^\alpha 
|k\rangle  \otimes \ e^{\tilde b_0^{\dagger} b_0^{\dagger} (\nu_k)}
|0\rangle \ , \label{g1} 
\ee
where we retained only the contribution from $y$
and $Z_0$. The normalization coefficients ${\cal N}_k$ can then
be fixed from the factorization of the tree-level 
amplitudes and are  given by 
\be
{\cal N}^b_0= T_p \sqrt {v_{p+1}} \, , \qquad 
{\cal N}^b_k= 2 \pi T_p \sqrt{v_{p-1} \over
\sin \pi \nu_k} \label{g2}
\ee
for D-branes and 
\be
{\cal N}^c_{k}= 2\pi \, T'_8 {1+(-1)^k \over 2}\, \sqrt{v_{8} \over  
\tan {\pi \nu_k \over 2}}  \label{g4}
\ee
for O8-planes, where $T'_8=16 T_8$ denote the tension of an O8 plane. 
Moreover, in the $\rho, \phi$ representation
one has  relation\footnote{In order to derive the last 
equality, we have used the
relation $\sum_{j=0}^{\infty} L_j^0 (x) t^j = (1/(1-t)) \exp [-{xt/(1-t)}]$.}
\be
\langle \rho,\phi | e^{\tilde b_0^{\dagger} b_0^{\dagger} (\nu_k)}
| 0 \rangle  = \sum_{j=0}^{\infty} \sqrt{\nu_k \over \pi \alpha'} \ 
(-1)^j L_j^0 (\nu_k \rho^2/\alpha') \ e^{- {\nu_k \rho^2 \over
2\alpha'}}
=  {\textstyle{1 \over 2}} \ \sqrt{\nu_k \over \pi\alpha'} \ . \label{g5}
\ee
It is remarkable that, despite the highly curved 
nature of the Melvin geometry, the boundary and crosscap states 
(\ref{g1}) are independent of $\rho$.

One can now extract, for example, the position of D-branes and O-planes
computing the scalar product 
\ba
f_b (y') &\equiv& 
\langle y' , \rho,\phi |b\rangle =T_p 
\sum_{k=0}^{\infty} \sqrt{{v_{p-1}\pi \nu_k \over \sin \pi\nu_k}}\, 
e^{{ik y' \over R'}} \, ,\\
f_c (y') &\equiv& 
\langle y' , \rho,\phi |c\rangle =T'_8 
\sum_{k=0}^{\infty} \sqrt{{v_{8}\pi \nu_{2k} \over \tan {\pi\nu_{2k}\over 2}}}\, 
e^{{i2k y' \over R'}} \,
\label{prof}
\ea
that presents some interesting novelties. 
Although in the limit of zero-twist
one recovers the familiar flat-space result $\langle y',\rho,\phi|
b\rangle \sim \delta ({ y'})$, $\langle y',\rho,\phi|c\rangle  \sim {1 \over 2} 
[\delta ({ y'})+\delta ({ y'}-\pi R')]$, corresponding to D8 branes located
at $y' =0$ and a pair of O8 planes sitting at $y' =0$ and $y' = \pi R'$,
for non-trivial $\gamma$ D-branes and O-planes are smeared on the $S^1$,
with a wave-function still centred around $y'=0,\pi R'$. The exact profile
can be easily found expanding  
$\sqrt{\pi\nu_k}\, {\cal N}^\alpha_k$ in a Fourier series.
Hence, the resulting profiles are
\be 
f_b (y') = {2\pi R' } \; \sum_m \hat {\cal F}^b_m \; \delta 
( y' + 2\pi m \gamma R' ) \,, \label{profile1}
\ee
for D-branes, and
\be
f_c (y' ) = {\pi R' } \; \sum_m \hat {\cal F}^c_m \;
\left[ \delta (y' + 2\pi m \gamma R' ) + \delta 
(y' - \pi R' + 2\pi m \gamma R' ) \right]\,, \label{profile2}
\ee
for O-planes, with
\ba
\hat {\cal F}^b_m &=& T_{p-1}\; \sqrt{v_{p-1}}\; \int _0^1 du\; \sqrt{\pi u 
\over \sin \pi u} \, e^{-2\pi i m u} \,, 
\nonumber \\
\hat {\cal F}^c_m &=& T'_8 \; \sqrt{v_8}\; \int _0^1 du\; \sqrt{\pi u 
\over \tan \pi u} \, e^{-2\pi  i m u} \,.
\ea
For large values of $m$, the integrals in right-hand side tend to
\be
 {1 \over 2\sqrt{|m|}}
\ (1- i \ {\rm sgn}(m)) \ .
\ee
As announced, O-planes and D-branes are distributed on the $S^1$ on a 
discrete, though dense for $\gamma$ irrational, set of points 
centred around $\pi R'$ and/or $0$, as in figure 1. 

\begin{figure}
\begin{center}
\epsfbox{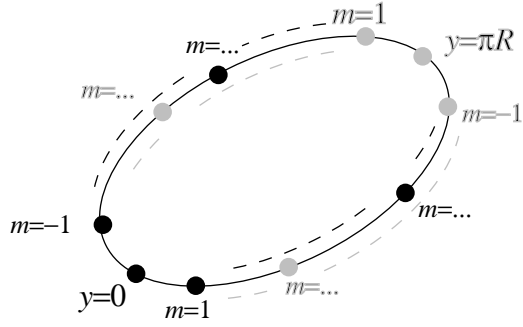}
\end{center}
\caption{Geometry of D8-branes and O8-planes. The black dots denote
D-branes and O-planes peaked around $y'=0$ and delocalized according to
(\ref{profile1}),(\ref{profile2}) . The grey dots denote O-planes peaked 
around $y'=\pi R'$ and delocalized according to (\ref{profile2}).}
\label{fig1}
\end{figure}

The results just obtained are straightforward to generalize for   
D-branes with Dirichlet boundary conditions in the non-compact $(\rho,
\phi)$ plane sitting at the origin of the Melvin plane, as discussed
for example in section 4.2 of \cite{dm}. The boundary state is in this
case 
\be
|b\rangle = \sum_{k} {\cal N}^b_k 
|k\rangle \otimes \ e^{- \tilde b_0^{\dagger} b_0^{\dagger} (\nu_k)}
|0\rangle \ , \label{g11} 
\ee
where the normalization coefficients
are
\be
{\cal N}^b_0 = T_p \sqrt{v_{p+1}} \,, \qquad 
{\cal N}^b_k= T_p \sqrt{\alpha' v_{p+1} \sin \pi \nu_k} \, . \label{g12}
\ee
The matrix element appearing in the boundary state for $\nu_k \not=0$ 
becomes here
\be
\langle \rho ,\phi | e^{-\tilde b_0^{\dagger} b_0^{\dagger} (\nu_k)}
| 0 \rangle = \sum_{j=0}^{\infty} \sqrt{\nu_k \over \alpha'\pi} \ 
L_j^0 (\nu_k \rho^2) \ e^{- {\nu_k \rho^2 \over 2}}
=   \delta (\rho^2)\sqrt{ \alpha' \over  {\pi \nu_k}} \ , \label{g13}
\ee
where in the last line we used the equality $\sum_{j=0}^{\infty}
 L_j^0 (x) = \delta(x)$.
The expression (\ref{g13}) shows unambiguously that the boundary state
is localized at the origin of the plane $\rho=0$. 
 
The profile of D-branes (and, similarly, of O-planes)
with Dirichlet boundary conditions in the non-compact $(\rho,\phi)$ 
plane is then given by
\be
f_b (y') \sim \sum_m \tilde{\cal F}^b_m \; \delta (y' + 2 \pi m \gamma R')
\,, \quad \hbox{\rm with now} \quad 
\tilde{\cal F}^b_m \sim \int _0^1 du\; \sqrt{\sin \pi u 
\over \pi u} \, e^{-2\pi i m u} \,.
\ee
Similarly to the previous cases  the profile is of
the form (\ref{profile1}) with $\hat {\cal F}$ now having the
asymptotic behaviour $\hat {\cal F} (m) \rightarrow 1 / (2\pi i m)$.

It is also rewarding to perform a classical analysis of the Kaluza-Klein 
expansion of closed-field couplings to O-planes. For instance, the 
wave-function for the excitation of the dilaton reads
\be
\Phi_{k,m,j}(\rho,\phi,{ y'}) =  \cos \left({k y' \over R'} \right)
\ e^{i m \phi} \ \sqrt{\omega_k^{|m|+1} j ! \over \pi \ 
(j+|m|)!} \ \ \rho^{|m|} L_j^{|m|} (\omega_k \rho^2) 
\ e^{- {\omega_k \rho^2 \over 2}} \ , \label{o8}
\ee
where, $m$ is an integer, $j$ is non-negative and labels the Landau levels, 
$\omega_k = |k| \gamma / \alpha'$, and the $L$'s 
are the Laguerre polynomials.
Following \cite{dm}, one can then compute the one-point functions of 
closed-string states with O-planes and D-branes. From the boundary state
(\ref{g1}) one finds the exact string-theory results
\be
g^2_{b,{\rm string}} (k,j) \equiv 
\langle k , j | b\rangle = T_p \sqrt{v_{p-1} \over \sin \pi \nu_k}\,,
\qquad 
g^2_{c,{\rm string}} (k,j) \equiv
\langle k , j | c\rangle = T_8 \sqrt{v_7 R' \over \tan \pi \nu_k}\,,
\ee
that, in general, differ from the semi-classical ones, obtained by
assuming a delta-function localized brane,
\be
{g^2_{b,{\rm classical}} (k,j) \over g^2_{b,{\rm string}} (k,j) } =
{\sin \pi \nu_k \over \pi \nu_k} \,, \qquad 
{g^2_{c,{\rm classical}} (k,j) \over g^2_{c,{\rm string}} (k,j) } =
{\tan \pi \nu_k \over \pi \nu_k} \, . 
\ee
They reduce to them for small $\gamma$, in analogy with D-branes in
WZW models \cite{bcw}. The novelty here is that the difference between
the string and the Born-Infeld results can be entirely attributed to
the delocalization of D-branes and O-planes on the $y'$ circle, 
as discussed above.
In fact, for $\gamma \ll 1$, it is possible to write the effective
low-energy coupling of closed-string states to branes:
\be
\int dy' d^{p+1}x \ \Phi(y',x) \ f_b(y') \ , \label{coup}
\ee
with $\Phi$ a function of closed-string modes.
Hence, taking (\ref{profile1}) into account turns eq. (\ref{coup}) into 
\be
2\pi R\sum_m \hat {\cal F}_{-m}^b \int d^{p+1}x \ \Phi (2\pi m\gamma R',x) 
\,,
\ee
suggesting that, on this Melvin background, a D-brane
develops an effective size in the $y'$ direction.
\section{Orientifolds of dual Melvin backgrounds}

We can now turn to the orientifold of the dual Melvin model in
eq. (\ref{buscher}). In this case, the world-sheet parity $\Omega$
is no longer a symmetry, and as such can not be used to project
the parent theory (\ref{tor3}). Orientifold constructions are, 
nevertheless, still possible if one combines the simple $\Omega$ with 
other symmetries as, for example, the parity $\Pi_\phi$ along $\phi$.
However, since the background (\ref{buscher}) is effectively curved,
it is simpler to construct such orientifolds starting from 
(\ref{c3}) and modding it out by the 
combination $\Omega' = \Omega \Pi_y \Pi_\phi$, with $\Pi_y$ 
a parity along $y$. Its action on the world-sheet coordinates 
$(y,Z_0 )$ is
\be
\Omega ' y (\sigma) \Omega ^{\prime\, -1}= - y (-\sigma) \,,
\quad
\Omega' Z_0 (\sigma ) \Omega ^{\prime\, -1} = Z_0^\dagger (-\sigma) \,,
\quad
\Omega' \lambda_+ (\sigma) \Omega ^{\prime\, -1} = \lambda^\dagger_- (-\sigma)
\,, \label{dmor}
\ee
that reflects in the action $\Omega ' J \Omega^{\prime\,-1} = - \tilde J$
on the angular momentum.
The angular momentum operators (\ref{angm}) reveal then that one set
of Landau levels of the IIB closed spectrum does propagate in the 
Klein bottle.

\begin{figure}
\begin{center}
\epsfbox{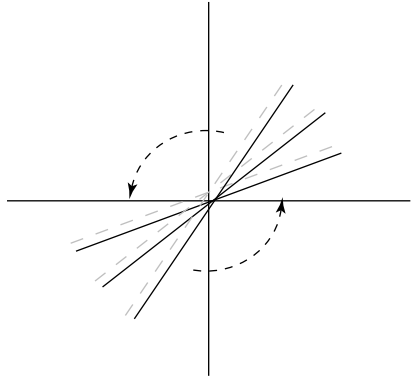}
\end{center}
\caption{Geometry of the O7-planes in the dual Melvin orientifold,
projection in the Z plane. The solid lines denote O7-planes and the
dashed lines denote anti O7-planes.}
\label{fig2}
\end{figure}

The orientifold identifications (\ref{dmor}) together with those in
eq. (\ref{one}) imply that the $y$ and $\phi$ coordinates satisfy
\be
y_0 = -y_0 + 2 \pi s R \,, \qquad \phi_0 = -\phi_0 + 2 \pi s \gamma +
2 \pi m \,,
\qquad (s,m \in \bb{Z})
\,, \label{d02}
\ee
whose fixed points
\be
(y_0 , \phi_0 )_1 = (s \pi R , s \pi \gamma)\,, \qquad
(y_0 , \phi_0 )_2 = (s \pi R , s \pi \gamma + \pi)\, ,
\label{fopl}
\ee
identify the positions of the O-planes. Actually, we are dealing now with 
two infinite sets of rotated O-planes, the angle being proportional to the
twist $\gamma$. Moreover, O-planes in different sets differing by an overall
$\pi$ rotation in the $\phi$ coordinate carry opposite R-R charge and,
as we shall see later on, the final configuration is neutral. 
Figures 2 and 3 give a pictorial representation of eq. (\ref{fopl}). 
\begin{figure}
\vspace{8 cm}
\special{hscale=60 vscale=60 voffset=0 hoffset=120
psfile=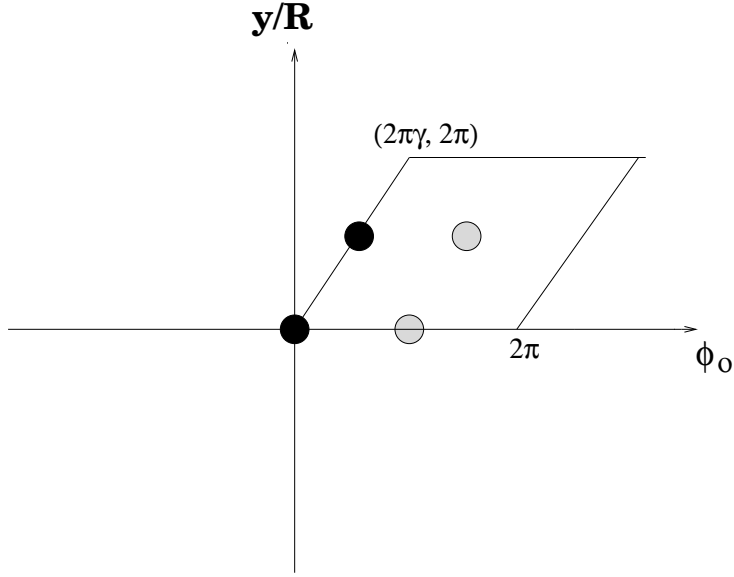}
\caption{Geometry of the O7-planes in the dual Melvin orientifold,
projection on the $y$-$\phi$ plane. The black circles denote O7-planes,
while the grey ones denote anti O7-planes.}
\label{fig3}
\end{figure}

The Klein-bottle amplitude corresponding to the projection (\ref{dmor})
of the model in (\ref{parti}) involves the Hamiltonian
\be
H = N - \nu J + {(n R)^2\over 4 \alpha'} + a \, , \label{d3}
\ee
and reads  
\be
{\cal K} = {\textstyle{1\over 2}} {v_7 \over \tau_2^{7/2} \eta^6}
\sum_n {\cal K} (n) \ q^{{1 \over 2 \alpha '} (nR)^2} \,,
\ee
with
\ba
{\cal K} (0) &=& {v_1 \over 2 \tau_2^{1/2} \eta^2} \sum_{\alpha , \beta}
\, {\textstyle{1\over 2}} \, \eta_{\alpha \beta} \, {\vartheta^4 \left[
{\alpha \atop \beta} \right] \over \eta^4} \,,
\nonumber \\
{\cal K} (n) &=&  \sum_{\alpha , \beta} \, {\textstyle{1\over 2}}\,
\eta_{\alpha \beta} \, e^{-\pi i  (2 \beta-1) \gamma n}\, 
{\vartheta^3 \left[ {\alpha \atop \beta}\right]
\over \eta^3}\, {\vartheta \left[ {\alpha + \gamma n \atop \beta}
\right] \over i \ \vartheta \left[ {1/2 + \gamma n \atop 1/2}
\right]} \,. \label{d4}
\ea
The first line in (\ref{d4}) contains a nontrivial numerical factor
${1\over 2}$, coming from integration over the non-compact momentum 
orthogonal to the O-planes, as described in (\ref{zerom1}). 
The map to the transverse channel presents similar problems to those
encountered in the direct-channel Melvin orientifolds. The amplitude
\be
\tilde{\cal K} = {2^{7/2} \over 2}\,{v_7 \over \eta^6} \, \ell^{1/2}\,
\sum_{\tilde k} \, \tilde{\cal K} (\tilde k) \, e^{-{\pi \over 2 \alpha' \ell}
({\tilde k} R)^2} \,, \label{pippo}
\ee
with
\ba
\tilde{\cal K} (0) &=& {2^{1/2} \, v_1 \over 2 \ell^{1/2}\, \eta^2} \,
\sum_{\alpha , \beta} {\textstyle{1\over 2}} \,\eta_{\alpha \beta}
{\vartheta^4 \left[{\alpha \atop \beta} \right] \over \eta^4} \,,
\nonumber \\
\tilde{\cal K} (\tilde k ) &=& -\sum_{\alpha , \beta} {\textstyle{1\over 2}}
\, \eta_{\alpha \beta} \, {\vartheta^3 \left[{\alpha \atop \beta} \right]
\over \eta^3}\, {\vartheta \left[ {\alpha \atop \beta + \gamma \tilde k}
\right] \over \vartheta \left[ {1/2 \atop 1/2 + \gamma \tilde k}
\right]} \,,
\ea
is not in a canonical form, since ${\tilde k}$ is {\it not} the KK
quantum number of the closed sector states. This makes the explicit 
evaluation of the tadpoles quite subtle. Nevertheless, from the structure of the 
amplitudes and from the $\Omega '$ action (\ref{fopl}) one can 
unambiguously deduce that pairs of O7 planes and anti-planes are present,
and, as a result, R-R tadpoles are not generated. In the following,
however, we shall resort to boundary states and quantum mechanical 
calculations to evaluate them.

Although the string partition function accounts for the whole
spectrum of physical states, built out from the vacuum with string
oscillators and/or KK modes, divergent contributions originate
only from massless states. Hence, to extract their tadpole it
suffices to restrict ourselves to the massless free Hamiltonian $H_0$ and
to its KK modifications
\ba
H &=& H_0 + {\alpha' \over 2} \left( {k \over R} - {\gamma \over R}  J
\right)^2 + a 
\nonumber \\
&=& {\alpha' \over 2} \left[ -{1 \over \rho} {\partial \over \partial \rho} 
\left(\rho {\partial
\over \partial \rho}\right) - {1 \over \rho^2} {\partial^2 \over \partial
\phi^2} - {\partial^2 \over \partial y^2} + {1 \over
R^2} \left( {k }+i\gamma {\partial
\over \partial \phi} \right)^2 \right] 
\,. \label{d6}
\ea
As a result, the leading contributions to the transverse-channel Klein-bottle
amplitude can be extracted from
\be
\langle c| e^{- \pi \ell H} |c\rangle 
= \sum_{\Psi} \langle c|\Psi\rangle  \ e^{-\pi \ell E_{\Psi}} \ 
\langle\Psi|c\rangle \ , \label{d5}
\ee
where $|\Psi\rangle$ are eigenstates of $H$ with
\be
\Psi_{E_\Psi ,k,m} (x_i,y) \equiv  \langle x_1,x_2,y|E_\Psi ,m,k\rangle  = 
A  e^{i{ky \over R}} \ e^{im \phi} \ J_{|m|}(\rho
\sqrt{E_\Psi -(k-\gamma m)^2/R^2}) \,, \label{d8}
\ee
where $A=({1 / 2 \pi \sqrt{2R}})$ is a normalization constant
and $E_\Psi \ge (k - \gamma m)^2 / R^2$ its  eigenvalues.
In (\ref{d8}), $J_p$ denote the Bessel functions of index $p$, $k$ is the
conventional KK momentum, and $m$ is the angular momentum conjugate to $\phi$.

Taking into account only the zero-mode contributions to the boundary
state $|c\rangle$ defined in the appendix, from
\be
\langle \Psi|c\rangle \sim  - {\cal N} [1+(-1)^k] [1+(-1)^m] \int_0^{\infty} d \rho
J_{|m|} (\rho a_{E,k,m}) = - {{\cal N} \over a_{E,k,m} } [1+(-1)^k] [1+(-1)^m] 
\,, \label{d10}
\ee 
where ${\cal N}$ is the normalization of the crosscap determined in the
Appendix, we arrive to the desired expression
\be
\langle c| e^{- \pi \ell H} |c\rangle_{{\rm NS-NS}} \sim 2 {\cal N}^2 \int_0^\infty 
{dE' \over 2E'} e^{- \pi \ell E'}
\sum_{m,k=-\infty}^{\infty} 
[1+(-1)^m]  [1+(-1)^k] \ e^{- {\pi \alpha' \ell \over 2R^2} ({k }-
\gamma m)^2} \,,
\label{d11}  
\ee
with $a_{E,k,m} \equiv \sqrt{E-(k-\gamma m)^2/R^2}$, and
$E' = E-(k - \gamma m)^2/R^2$.
{}From (\ref{d11}) we can then extract the non-vanishing dilaton tadpole
(for $m=0$ and $k=0$), as well as informations about the geometry of
the O-planes. The factor $1+(-1)^k$ suggests that in the $y$ direction
there are O-planes sitting at $0$ and $\pi R$, while the projector
$1+(-1)^m$ implies that in the $(\rho,\phi)$ space there are pairs
of orientifold planes rotated by a $\pi$ angle, in agreement with 
(\ref{fopl}). 

Following the same procedure, we can now extract the R-R tadpoles.
For R-R states the angular momentum along the two-plane
has components $J=\tilde J = \pm {1\over 2}$, and, as a result,
the mass is shifted by one
\ba
M^2 |c, m, k=0 \rangle &=& {\gamma^2 \over R^2} \ 
(m- \zeta_0^{\dagger} \zeta_0+ 
{\tilde \zeta}_0^{\dagger} {\tilde \zeta}_0 )^2 \ |c, m, k=0\rangle 
\nonumber \\
&=& {\gamma^2 \over R^2} (m \mp 1)^2 |c, m, k=0\rangle 
\,. \label{d12}
\ea
The massless
R-R states, which define the R-R charge of the orientifolds, have
therefore orbital angular momentum $m=\pm 1$. The R-R portion
of the amplitude is
\ba
\langle c| e^{- \pi \ell H} |c\rangle_{{\rm R-R}} &\sim & 2 {\cal N}^2 
\int_0^\infty 
{dE' \over 2E'} e^{- \pi \ell E'}
\sum_{m,k=-\infty}^{\infty} 
[1+(-1)^m]  [1+(-1)^k] \ e^{- {\pi \alpha' \ell \over 2R^2} ({k }-
\gamma (m\mp 1))^2} 
\nonumber \\
&=& 2 {\cal N}^2 \int_0^\infty 
{dE' \over 2E'} e^{- \pi \ell E'}
\sum_{m',k=-\infty}^{\infty} 
[1-(-1)^{m'}]  [1+(-1)^k] \ e^{- {\pi \alpha' \ell \over 2R^2} ({k }-
\gamma m')^2} \,,
\label{d11rr}  
\ea
has no contributions for massless R-R states, and nicely encodes 
the geometry (\ref{fopl}) 
of the orientifold planes: from 
\be
[1-(-1)^{m'}]  [1+(-1)^k] = 1 -(-1)^{m'} + (-1)^k - (-1)^{m' + k}
\ee
one can read that O-planes are located at $(y_0 , \phi_0)_1$
while O-antiplanes are sitting at $(y_0 , \phi_0)_2$.
Up to the presence of images, this
phenomenon of the occurrence of O-$\overline{\rm O}$ systems is similar to the
one encountered in \cite{ads1} and geometrically interpreted in
\cite{ad}.

Let us now turn to the open sector of the orientifold. 
Actually in this case one is not demanded to add D-branes 
since, as we have seen, O-planes yield a vanishing R-R tadpole.
Nevertheless we shall introduce brane-antibrane pairs to compensate
(globally and locally) the tension of the orientifold planes and
preserve the structure of the vacuum.  

The open string Hamiltonian
\be
H = N - \nu J + { (n  R)^2 \over \alpha'} + a \, ,
 \label{d191}
\ee
with now $\nu=f(2n\gamma)$, determines the direct-channel 
annulus 
\be
{\cal A} = {\textstyle{1\over 2}} {v_7 \over \tau_2^{7/2} \, \eta^6}
\sum_n \left[ (n_+^2 + n_-^2 ) {\cal A}_1 (n) q^{{1\over 2\alpha'} 
(nR )^2} + 2 n_+ n_- {\cal A}_2 (n) q^{{1\over 2\alpha '} [(n+{1\over 2})R]^2}
 \right]\,, \label{d0191}
\ee
with
\ba
{\cal A}_i (0) &=& {v_1 \over \tau_2^{1/2} \eta^2} \sum_{\alpha,\beta}
\, {\textstyle{1\over 2}}\, \eta^i_{\alpha\beta}\, {\vartheta^4 \left[
{\alpha \atop \beta} \right] \over \eta^4} \,,
\nonumber \\
{\cal A}_i (n) &=&  \sum_{\alpha,\beta}\, {\textstyle{1\over 2}} \,
\eta^i_{\alpha\beta} e^{- 2 \pi i \gamma n (2 \beta-1)} \, 
{\vartheta^3 \left[ {\alpha \atop \beta}
\right] \over \eta^3} \, {\vartheta \left[ {\alpha + 2 \gamma n \atop
\beta} \right] \over i \ \vartheta \left[ {1/2 + 2 \gamma n \atop
1/2} \right] } \,,
\ea
and M{\"o}bius amplitudes
\be
{\cal M} = -{\textstyle{1\over 2}} {v_7 \over \tau_2^{7/2}\, \eta^6} 
\sum_n (n_+ + n_-) \left[ {\cal M}_{\rm NS} (n) + (-1)^n {\cal M}_{\rm R}
\right] q^{{1\over 2\alpha'} (nR)^2}
\,,
\ee
with
\ba
{\cal M}_{\rm NS,R} (0) &=& {v_1 \over \tau_2^{1/2} \eta^2} \sum_{\beta}
\, {\textstyle{1\over 2}}\, \eta^1_{\alpha \beta}\, {\vartheta^4 \left[
{\alpha \atop \beta} \right] \over \eta^4} \,,
\nonumber \\
{\cal M}_{\rm NS,R} (n) &=&  \sum_{\beta}\, {\textstyle{1\over 2}} \,
\eta^1_{\alpha \beta} e^{- 2 \pi i \gamma n (2 \beta -1)} 
\, {\vartheta^3 \left[ {\alpha \atop \beta}
\right] \over \eta^3} \, {\vartheta \left[ {\alpha + 2 \gamma n \atop
\beta - \gamma n} \right] \over i \ \vartheta \left[ {1/2 + 2 \gamma n \atop
1/2 - \gamma n} \right] } \,.
\ea
Here $\eta^i_{\alpha,\beta}$ correspond to the two 
complementary GSO projections, $\eta^1_{\alpha \beta} = \eta_{\alpha \beta}$,
$\eta^2_{\alpha \beta} = (-1)^{2\alpha + 4 \alpha \beta}$, 
while in ${\cal M}$ $\alpha =0$ ($\alpha ={1\over 2}$)
in the R (NS) sector. The Chan--Paton multiplicities $n_\pm$ count
the number of branes and antibranes, while, as usual, the theta and eta
functions depend on the modulus of the doubly-covering torus: ${1\over 2} i
\tau_2$ for the annulus and ${1\over 2} + {1\over 2} i \tau_2$ for the
M{\"o}bius. Notice the dependence of ${\cal A}$ and ${\cal M}$ on a double
twist $2 \gamma n$ in the upper characteristic. It is due to the horizontal
doubling of the corresponding elementary cells, and is similar in spirit
to the doubling of the lower characteristic of ${\cal K}$ in eqs. 
(\ref{kdirm}) and (\ref{kdirmb}). For irrational $\gamma$
the gauge group is ${\rm SO}(N)$. The global NS-NS tadpole condition 
\be
\langle \Psi | c \rangle +\langle \Psi | b \rangle = 0 \ ,  
\ee
is solved, by using the Appendix, by computing
\be
\langle \Psi|b\rangle \sim   (n_{+}+n_{-}) 
{{\cal N} \over 8}  [1+(-1)^m] \int_0^{\infty} d \rho
J_{|m|} (\rho a_{E,k,m}) =  {{\cal N} \over 8 a_{E,k,m} } [1+(-1)^m] 
\,, \label{d010}
\ee 
and fixes therefore $n_{+}+n_{-}=16$. The low-lying open-string excitations is not
affected by the twist $\gamma$ and thus one would conclude that the
massless D-brane spectrum is supersymmetric. However, for $\gamma >
2R^2/\alpha'$ tachyons can appear in the antisymmetric representation.
Alternatively, the model has a gauge group ${\rm SO}(n_+)
\otimes {\rm SO}(n_-)$, with $n_+ = n_- = 8 $ if we want to
cancel locally the NS-NS tadpole. The local cancellation conditions must
be satisfied whenever the Melvin model (\ref{one}) is embedded into
M-theory, with $y$  identified with the eleventh coordinate.
 
Finally, it is quite rewarding to study the limit $\gamma \to 1$,
that, as expected, reproduces the M-theory 
breaking of \cite{ads1}. Actually, this limit presents some further
subtleties that are not only associated to a proper accounting of the
zero-modes. Although for generic $\gamma$ $\Omega'$-invariant
configurations of D-branes share with the O-planes the property of having
zero R-R charge, for $\gamma=1$ KK states can compensate the mass-shift
due to the internal angular momentum and a non-vanishing charge is
generated. To be more specific, the leading R-R coupling for D-branes 
reads
\be
\langle b | e^{-\pi\ell H} |b\rangle_{\rm R-R} \sim 2 \int_0^\infty
{dE' \over 2 E'} \, e^{-\pi\ell E'} \sum_{m,k} \left[
1 + (-1)^m \right] \, e^{-{\pi\alpha'\ell \over 2 R^2} [ k - \gamma (m
\mp 1)]^2} \,,
\ee
where all, unconstrained, KK states contribute to it. Hence, when 
$\gamma = 1$ the states with $k = \pm 1$ become massless and carry a
definite R-R charge. This has to be contrasted to the situation for O-planes
where the projector on even $k$ does not afford this possibility.

\section{Tachyons in Melvin backgrounds}

Let us add now few remarks about closed- and open-string tachyons in
these orientifold Melvin backgrounds. As we have anticipated in section 2, the 
Melvin model (\ref{tor2}) has complex tachyons
\ba
&& \mu^\dagger_1\tilde \mu_1^\dagger|k=0,n=1\rangle \,, \qquad 
\zeta_0^\dagger\tilde \zeta_0^\dagger|k=0,n=-1\rangle  \,, \qquad 
({\rm for}\ \gamma<{\textstyle{1\over 2}}) \, ,
\nonumber \\
&& |k=0,n=1\rangle \,, \qquad |k=0,n=-1\rangle \,, \qquad ({\rm for}\ 
{\textstyle{1 \over 2}} <\gamma<1) \,, \label{t1}
\ea
whenever $R^2 < 2\alpha' \gamma $, whose mass is given by (\ref{tmass}). 
These are not independently 
invariant under the world-sheet parity
$\Omega$, though their linear combination is. As a result, the unoriented 
closed-string spectrum comprises a real tachyon that, due to a change in
GSO projections, can be identified either with
\be
\mu_1^\dagger \tilde\mu_1^\dagger |k=0 , n=1\rangle +
\zeta_0^\dagger \tilde\zeta_0^\dagger |k=0 , n=-1 \rangle
\ee
if $\gamma < {1\over 2}$, or with a linear combination of the NS-NS vacuum
with one unit of winding number if $\gamma > {1\over 2}$. This real tachyon
propagates in $\tilde{\cal K}$, $\tilde{\cal A}$ and $\tilde{\cal M}$ and thus
couples to both O-planes and D-branes. Furthermore, the D-brane spectrum 
is free of open-string tachyons, and it would be tempting to speculate
that this vacuum configuration decays into the SO(32) superstring.

Quite different is the dual Melvin case. The parent model 
(\ref{tor3})-(\ref{parti2}) share with (\ref{tor2})-(\ref{parti}) 
the same tachyons for $R^2 < 2 \alpha ' \gamma$
but now both are invariant under the modified world-sheet parity
$\Omega '$, and hence, the unoriented closed-string spectrum includes a 
complex tachyon. Moreover, since the model involves pairs of (image)
branes and antibranes, open-string tachyons are present as well, though
only for $R^2 < {1\over 2} \alpha ' \gamma$. A further difference with 
the previous case comes from the transverse-channel amplitudes: the
closed-string tachyons do not propagate and thus do not couple neither
to O-planes nor to D-branes. 

To conclude this very brief analysis of tachyons in Melvin backgrounds, let 
us recall that for $\gamma =1$ the dual Melvin model affords a second
projection $\Omega ''
= \Omega \Pi_y (-1)^{f_{\rm L}}$, with $f_{\rm L}$ the left world-sheet 
fermion number. This is a natural extension of the non-tachyonic 0B 
orientifolds first introduced in \cite{augusto} and studied in this context in
\cite{dm3}, and has the virtue of eliminating the closed-string tachyons for
any value of the radius, since the lowest mass scalar has the KK and winding number
$|k=0,n=1\rangle $. Its mass 
\be
M_T^2 = -{2 \over \alpha'} + {1 \over R^2} + {R^2 \over \alpha'^2} =  
\left({1 \over R}- {R \over \alpha'} \right)^2   \label{t2}
\ee
is positive and becomes zero at the self-dual value for the radius.
The fate of this 
orientifold is an interesting and open problem, since it is
tachyon-free. Non-perturbative instabilities of the type studied in
\cite{witten} can still occur and a more detailed analysis would be
of clear interest.

\section{Double Melvin backgrounds and supersymmetry restoration}

Until now we have considered the case of a single magnetized two-plane,
whose complex coordinate is actually twisted. We have also seen how
their orientifolds have quite interesting properties. Much more appealing
and surprising features emerge if we consider the case of multiple two-planes
subject to independent twists. In this case we have the option to couple
each twist to the same $S^1$ or to different ones. While the latter 
corresponds to a trivial generalization of the models previously studied
and, as such, shares with them all their salient properties, the former
turns out to be quite interesting and leads to amusing phenomena.

To be more specific let us consider the simple case of two two-planes, labelled
by coordinates $(\rho_1 , \phi_1)$ and $(\rho_2 , \phi_2)$, coupled to the
same circle of radius $R$ parametrized by the compact coordinate $y$. 
The world-sheet Action then reads
\ba
S &=& - {1 \over 4 \pi \alpha '} \int \left[ d\rho_1 \wedge ^* d \rho_1
+ \rho_1^2 \left( d\phi_1 + {\gamma_1 \over R} dy\right) \wedge ^*
\left( d\phi_1 + {\gamma_1 \over R} dy \right) \right.
\label{doubsm} \\
& & + \left. d\rho_2 \wedge ^* d \rho_2
+ \rho_2^2 \left( d\phi_2 + {\gamma_2 \over R} dy\right) \wedge ^*
\left( d\phi_2 + {\gamma_2 \over R} dy \right) + dy \wedge ^* dy \right] \,,
\nonumber
\ea
where $\gamma_1$ and $\gamma_2$ are the two twists.

The quantization procedure is a simple generalization of the single-twist
case previously studied and yields the torus partition function
\be
{\cal T}_1 = \sqrt{R^2 \over \alpha ' \tau_2} {v_5 \over \tau^{5/2}_2 |\eta|^8}
\sum_{\tilde k , n}  {\cal T}_1 (\tilde k , n ) \, 
e^{-{\pi R^2 \over \alpha ' \tau_2} |\tilde k + \tau n |^2} \,, \label{dmt}
\ee
with
\be
{\cal T}_1 (0,0) = {v_4 \over \tau_2^2 |\eta|^8} \left| \sum_{\alpha ,\beta}
{\textstyle{1\over 2}} \, \eta_{\alpha \beta} \,
{\vartheta^4 {\textstyle{\left[ 
\alpha \atop \beta \right]}} \over \eta^4} \right|^4
\ee
and
\be
{\cal T}_1 
(\tilde k , n ) = \left| \sum_{\alpha , \beta} {\textstyle{1\over 2}} \,
\eta_{\alpha \beta} \, {\vartheta^2 {\textstyle{\left[ \alpha 
\atop \beta \right]}}
\over \eta^2} \prod_{i=1,2} e^{-2 \pi i \beta \gamma_i n} 
{\vartheta {\textstyle {\left[ \alpha + \gamma_i n 
\atop \beta + \gamma_i \tilde k \right]}} \over 
\vartheta {\textstyle {\left[ 1/2 + \gamma_i n 
\atop 1/2 + \gamma_i \tilde k \right]}}} \right|^2 \,.
\ee

Also in this case one can generate a new interesting background performing 
a Buscher duality along the $y$ coordinate. This results in the interchange
of windings and momenta and leads to the alternative partition function
\be
{\cal T}_2 = \sqrt{\alpha ' \over R^2 \tau_2} {v_5 \over \tau_2^{5/2}
|\eta|^8} \sum_{k,\tilde n} {\cal T}_2 (k,\tilde n) \ e^{- {\pi \alpha' \over R^2 \tau_2} 
|\tilde n + \tau k|^2} \,, \label{ddmt}
\ee
with
\be
{\cal T}_2 (0,0) = {v_4 \over \tau_2^2 |\eta|^8} \left| \sum_{\alpha , \beta}
{\textstyle{1\over 2}}\, \eta_{\alpha \beta} \,
{\vartheta^4 {\textstyle{ \left[
\alpha \atop \beta \right]}} \over \eta^4 } \right|^2
\ee
and
\be
{\cal T}_2 (k, \tilde n ) = \left| \sum_{\alpha , \beta} {\textstyle{1\over 2}}
\, \eta_{\alpha \beta} \,
{\vartheta^2 {\textstyle{\left[ \alpha \atop \beta \right]}}
\over \eta^2} \prod_{i=1,2}  e^{-2 \pi i \beta \gamma_i k}
{\vartheta {\textstyle{\left[ \alpha + \gamma_i k 
\atop \beta + \gamma_i \tilde n \right]}} 
\over \vartheta {\textstyle{\left[ 1/2 + \gamma_i k 
\atop 1/2 + \gamma_i \tilde n \right]}}}  \right|^2 \ . 
\ee

A careful reading of these amplitudes reveals that something special 
happens if $\gamma_1 = \pm \gamma_2$. The partition functions vanish 
identically, a suggestive signal of supersymmetry restoration. Indeed,
a simple analysis of Killing spinors for the background (\ref{doubsm}) 
shows that whenever $\gamma_i$ are even integers all supersymmetry charges
are preserved \cite{Russo:2001na}. However,
one has now the additional possibility of preserving only half of the 
original supersymmetries if
\be
\gamma_1 \pm \gamma_2 \in 2 \bb{Z} \ . 
\ee
This is very reminiscent of the condition one gets for orbifold 
compactifications. Indeed, if the two two-planes were compact and, as a 
result, the $\gamma_i$ had to meet some quantization conditions, one would 
find the familiar result $\gamma_1 = \pm \gamma_2$ in order to have a 
supersymmetric spectrum.

Particularly interesting is the case $\gamma_i = {1 \over 2}$. After a 
careful handling of the zero-mode contributions, the partition 
functions take a simple form and, actually, reproduce the Scherk-Schwarz
partial supersymmetry breaking of 
\cite{ssopen}. After a proper redefinition of the 
radius, the resulting amplitudes correspond to the orbifold $(\bb{C}^2 \times
S^1 ) / \bb{Z}_2$, where the $\bb{Z}_2$ acts as a reflection on the 
$\bb{C}^2$ coordinates accompanied by a momentum or winding shift along
the compact circle. As a result, one can use the $\gamma$'s to smoothly 
interpolate among ${\cal N} = 8$ vacua, ${\cal N} = 4$ vacua and ${\cal N}
=0 $ vacua, all in the same space-time dimensions\footnote{We remind the 
reader that in standard Scherk-Schwarz compactifications the restoration
of maximal (super)symmetries corresponds to a decompactification limit.}.
Given ${\cal T}_1$ and ${\cal T}_2$ and the results in the previous
sections we can now proceed to compute their orientifolds. 

\section{Orientifolds of double-Melvin backgrounds}

Let us start considering the $\Omega$ projection of the more conventional 
amplitude ${\cal T}_1$. As in standard orientifold constructions the 
Klein-bottle amplitude receives contributions from those states that are
fixed under $\Omega$. These are nothing but the strings with vanishing
winding numbers. As a result the amplitude in the direct channel reads
\be
{\cal K} = {\textstyle{1\over 2}} {R \over \sqrt{\alpha ' \tau_2}} \,
{v_5 \over \tau_2^{5/2} \, \eta^4} \sum_{\tilde k} {\cal K} (\tilde k)
\, e^{- {\pi \over \alpha '  \tau_2} (\tilde k R )^2}  \,,
\ee
where
\be
{\cal K} (0) = {v_4 \over \tau_2^2 \, \eta^4 } \sum_{\alpha , \beta}
{\textstyle{1\over 2}} \eta_{\alpha \beta} {\vartheta^4 {\textstyle{
\left[ \alpha \atop \beta \right]}} \over \eta^4} (2 i \tau_2)
\ee
and
\be
{\cal K} (\tilde k) = \sum_{\alpha \beta} {\textstyle{1\over 2}}\, 
\eta_{\alpha \beta} \,
{\vartheta^2 {\textstyle{\left[ \alpha \atop \beta \right]}} \over \eta^2}
\prod_{i=1,2} {2 \sin (2 \pi \gamma_i \tilde k ) \over \left[ 2 \sin 
(\pi \gamma_i \tilde k ) \right]^2} {\vartheta {\textstyle{\left[ \alpha \atop
\beta + 2 \gamma_i \tilde k \right]}} \over
\vartheta {\textstyle{\left[ 1/2 \atop
1/2 + 2 \gamma_i \tilde k \right]}}  } (2 i \tau_2) \,.
\ee

In writing these amplitudes we have taken into account that they depend on
the modulus of the doubly-covering torus, that is obtained by a vertical
doubling. Hence, the twist in the temporal direction is effectively doubled
and, as a result, the lower characteristic depends on $2 \gamma_i \tilde k$.
On the contrary, the contribution of the zero modes is not affected.

As in the single-twist case, it is hard to extract any information from this
amplitude, given the unconventional $\tau_2$-dependence of the lattice 
contribution. The transverse-channel amplitude, however, has the standard
structure
\be
\tilde{\cal K} = {2^3 \over 2} {R \over \sqrt{\alpha '}} \, {v_5 \, \ell 
\over \eta^4} \sum_n \tilde {\cal K} (n) \, q^{{1\over \alpha '} (nR)^2} \,,
\ee
with
\be
\tilde{\cal K} (0) = {2^2 \, v_4 \over \eta^4} \sum_{\alpha , \beta}
{\textstyle{1\over 2}} \, \eta_{\alpha \beta} \, {\vartheta^4 {\textstyle{
\left[ \alpha \atop \beta \right]}} \over \eta^4 } (i\ell )
\ee
and
\be
\tilde{\cal K} (n) = \sum_{\alpha ,\beta} {\textstyle{1\over 2}} \,
\eta_{\alpha \beta} \, 
{\vartheta^2 {\textstyle{\left[ \alpha \atop \beta \right]}}
\over \eta^2} \prod_{i=1,2} e^{-2 \pi i \gamma_i n (2 \beta-1)} 
{2 \sin (2 \pi \gamma_i n) \over
\left[ 2 \sin (\pi \gamma_i n ) \right]^2} 
{\vartheta {\textstyle{\left[ \alpha + 2 \gamma_i n
\atop \beta \right]}} \over i \ \vartheta {\textstyle{\left[ 1/2 + 2 \gamma_i n
\atop 1/2 \right]}}} (i\ell) \,,
\ee
and develops non-vanishing tadpoles. From these amplitudes we can also extract
interesting informations about the couplings of closed-string fields to
orientifold planes and their geometry.

Before turning to the open-string sector, it is interesting to give a 
closer look at ${\cal K}$ and take the limit $\gamma_i \to {1\over 2}$.
We get 
\be
{\cal K} = {\textstyle{1\over 4}} \, {v_5 \over \tau_2^{5/2} \, \eta^4} \sum_k
{\cal K} (0) \, \left[ 1 + {\tau_2^2 \over v_4} (-1)^k \right] \,
q^{{\alpha ' \over 2} \left( {k \over R} \right)^2}\,.
\ee
This amplitude and its transverse-channel counterpart
\be
\tilde{\cal K} = {2^3 \over 2} \sqrt{R^2 \over \alpha '} {v_5 \over \eta^4}
\sum_n \tilde{\cal K} (0) \, \left[ q^{{1\over 4\alpha'} (4nR)^2} 
+ {\ell^2 \over v_4} \, q^{{1\over 4\alpha'} (4(n+1/2)R)^2} \right] 
\ee
immediately spells out the geometry of the O-planes: one has standard O9 planes
that invade the whole space-time and contribute to NS-NS and R-R tadpoles,
and a pair of O5 planes located at $y=0$ and $y=\pi R$ with opposite
tension and R-R charge. The presence of O5 planes is encoded in the
$\tau_2^2 / v_4$ term that exactly cancels a similar one hidden in 
$\tilde {\cal K} (0)$. Furthermore, their relative tension and charge can 
be extracted, as usual, from the corresponding term in $\tilde{\cal K}$ that 
involves only odd windings.

We can now turn to the open-string sector and, in particular, to the 
transverse-channel annulus amplitude
\be
\tilde {\cal A} = {2^{-3} \over 2} \, {R \over \sqrt{\alpha '}} 
{v_5 \, \ell \over \eta^4} 
\sum_n  N_n^2 \, \tilde{\cal A} (n)\, q^{{1\over 4\alpha '} (nR)^2} \,,
\ee
with
\be
\tilde{\cal A} (0) = {2^{-2} \, v_4 \over \eta^4} \sum_{\alpha , \beta}
{\textstyle{1\over 2}} \, \eta_{\alpha \beta}\, 
{\vartheta^4 {\textstyle{\left[
\alpha \atop \beta \right]}} \over \eta^4} (i\ell )
\ee
and
\be
\tilde{\cal A} (n) = \sum_{\alpha , \beta} {\textstyle{1\over 2}} \,
\eta_{\alpha \beta} \, {\vartheta^2 {\textstyle{\left[ 
\alpha \atop \beta \right]}}
\over \eta^2} \prod_{i=1,2}  e^{- \pi i \gamma_i n (2 \beta-1)}
{1\over 2 \sin (\pi \gamma_i n)}
{\vartheta {\textstyle{\left[ \alpha + \gamma_i n \atop \beta \right]}} \over
i \ \vartheta {\textstyle{\left[ 1/2 + \gamma_i n \atop 1/2 \right]}} } (i\ell ) 
\,.
\ee

The transverse-channel M{\"o}bius amplitude is then entirely determined 
from $\tilde{\cal K}$ and $\tilde{\cal A}$ and reads
\be
\tilde{\cal M} = - {R \over \sqrt{\alpha '}} {v_5 \, \ell \over \eta^4}
\sum_n N_{2n}\, \tilde {\cal M} (n) \, q^{{1\over 4\alpha'} (2nR)^2}  \,,
\ee
with
\be
\tilde{\cal M} (0) = {v_4 \over \eta^4} \, \sum_{\alpha , \beta} 
{\textstyle{1\over 2}} \, \eta_{\alpha \beta}\, 
{\vartheta^4 {\textstyle{ \left[
\alpha \atop \beta \right]}} \over \eta^4 } (i\ell + {\textstyle{1 \over 2}})
\ee
and
\be
\tilde{\cal M} (n) = \sum_{\alpha , \beta} {\textstyle{1\over 2}} \,
\eta_{\alpha \beta} \,
{\vartheta^2 {\textstyle{\left[ \alpha \atop \beta\right]}} \over
\eta^2} \prod_{i=1,2} e^{-2 \pi i \gamma_i n (2 \beta-1)}
{1 \over 2\sin (\pi\gamma_i n)}  {\vartheta {\textstyle{
\left[\alpha + 2 \gamma_i n \atop \beta - \gamma_i n \right]}} \over
i \ \vartheta {\textstyle{
\left[1/2 + 2 \gamma_i n \atop 1/2 - \gamma_i n \right]}} } 
(i\ell + {\textstyle{1 \over 2}}) \, .
\ee
The global tadpole conditions fixes the gauge group to be SO(32) or
a Wilson line breaking of it, while if we insist in imposing the 
local tadpole conditions we
find $N_{2n} = 32 \ {\rm and} \ N_{2n+1} = 0$, with a gauge group
${\rm SO}(16) \times {\rm SO}(16)$.
{}From these amplitudes one can then extract the one-point couplings
of closed-string states in front of boundaries as well as the geometry
of the branes, that are an obvious generalization of those in section 5.

Also for this orientifold one can consider the interesting particular
case $\gamma_i = {1\over 2}$.
The annulus and M{\"o}bius amplitudes describe then the propagation of D9 
branes only, and nicely reproduce the non-compact versions of the
results obtained in \cite{ssopen}.

\section{Dual double-Melvin orientifolds}

To conclude we can now turn to analyse orientifolds of the dual 
double-Melvin background. As in section 5, $\Omega$ is not a symmetry
of the IIB model (\ref{ddmt}), rather we should combine it with parity
transformations on the two angular variables. However, since the 
background corresponding to eq. (\ref{ddmt}) is actually curved, it is
simpler to study the $\Omega ' = 
\Omega \Pi_y \Pi_{\phi_1} \Pi_{\phi_2}$ orientifold
of the model (\ref{dmt}), that yields similar results. $\Omega '$ has fixed
points at
\ba
&& (y_0 , \phi_{0,1} , \phi_{0,2} )_1 = (s \pi R , s \pi \gamma_1
, s\pi\gamma_2) \ , \nonumber \\
&& (y_0 , \phi_{0,1} , \phi_{0,2} )_2 =  (s \pi R , 
s \pi \gamma_1 + \pi , s \pi \gamma_2 + \pi )\,,
\nonumber \\
&& (y_0 , \phi_{0,1} , \phi_{0,2} )_3 = (s \pi R  , s \pi \gamma_1 ,
s \pi \gamma_2 + \pi ) \ , \nonumber \\
&&(y_0 , \phi_{0,1} , \phi_{0,2} )_4 = (s \pi R , s \pi \gamma_1+ \pi 
 , s \pi \gamma_2  ) \ , 
\label{ddmfopl}
\ea
that, as usual, accommodate rotated O-planes. 
For the generic non-supersymmetric case $\gamma_1 \not= \gamma_2$,
similar arguments to that presented in section 5 show that the system
has zero R-R charge and therefore orientifold planes have images with
total vanishing R-R charge.  

For the supersymmetric case $\gamma_1 = \pm \gamma_2$, however, 
as contrasted to the case
of a single twist, there are no anti-O-planes in this double-Melvin
orientifold. Actually, a special case is $\gamma_1 = r \gamma_2$, where
$r$ is an integer. In this case, there are massless RR states which 
do couple to
charged O-planes, which suggests that anti O-plane images are not
generated in this case.
Furthermore, as anticipated
from our previous discussions and as we shall see in the following,
something special happens for $\gamma_1 = \gamma_2 = {1\over 2}$: 
mutually orthogonal O6 planes are generated. Indeed, points with $s$ and
$s+1$ in (\ref{ddmfopl}) are mutually rotated by $\pi/2$ in two
complex planes, as pertains to a 
(T-dualized) $(\bb{C}^2 \times S^1)/\bb{Z}_2$ orientifold. Actually,
the resulting configuration is more involved and we shall return 
shortly on this point.

The Klein-bottle amplitude
\be
{\cal K} = {\textstyle{1\over 2}} \, {v_5 \over \tau_2^{5/2} \, \eta^4}
\sum_n {\cal K} (n) \, q^{{1\over 2\alpha '} (nR)^2} \,, \label{ddmk}
\ee
with\footnote{As for the simpler model described in section 5, a non-trivial
factor of ${1\over 4}$ in ${\cal K} (0)$ appears as a consequence of
integrating over the two non-compact momenta orthogonal to the O6 planes.}
\ba
{\cal K} (0) &=& {v_2 \over 4 \tau_2 \, \eta^4} \sum_{\alpha , \beta}
{\textstyle{1\over 2}} \, \eta_{\alpha \beta} \, {\vartheta^4 \left[
{\alpha \atop \beta} \right] \over \eta^4} \,,
\nonumber \\
{\cal K} (n) &=& \sum_{\alpha , \beta} {\textstyle{1\over 2}} \, 
\eta_{\alpha \beta} \, {\vartheta^2 \left[ {\alpha \atop \beta} \right] 
\over \eta^2} \, \prod_{i=1,2} { \vartheta \left[ {\alpha + \gamma_i n
\atop \beta} \right] \over i \ \vartheta \left[ {1/2 + \gamma_i n \atop
1/2}\right]} \, e^{-i\pi(2 \beta -1) \gamma_i n} \,,
\ea
clearly spells out the geometry of the O6 planes, and, 
in the limit $\gamma_i \to {1\over 2}$, becomes
\be
{\cal K} = {\textstyle{1\over 2}} \, {v_5 \over \tau_2^{5/2}\, \eta^4} 
\sum_n
\left[ {\cal K} (0) \, q^{{\alpha ' \over 2} ( 2 n R)^2} + 
{\cal K}_{\rm t} (0) \, q^{{\alpha ' \over 2} 
[( 2 n +1 )R ]^2} \right] \,,
\ee
where now
\be
{\cal K}_{\rm t} (0)= \sum_{\alpha , \beta} {\textstyle{1\over 2}} \,
(-1)^{2\alpha + 4 \alpha \beta} \, {\vartheta^2 \left[ {\alpha \atop
\beta} \right] \over \eta^2}\, {\vartheta^2 \left[ {\alpha + 1/2 \atop
\beta} \right] \over \vartheta^2 \left[ {0\atop 1/2} \right]} \,.
\ee
According to our previous analysis, the emergence of ${\cal K}_{\rm t}$
implies the presence of mutually orthogonal O6-planes that, indeed, 
form a BPS configuration preserving one quarter of supersymmetries of
the parent closed-string model. In fact, in this limit, eq. (\ref{ddmk})
reproduces the ${\cal N} = 4 \to {\cal N} = 2$ M-theory breaking of
\cite{ssopen}.

We can now turn to the open-string sector where, now, tadpole cancellation
would require the introduction of two types ($N$ and $D$ in the following)
of rotated branes. The annulus and M{\"o}bius amplitudes thus read
\be
{\cal A} = {\textstyle{1\over 2}} {v_5 \over \tau_2^{5/2}\, \eta^4}
\sum_n \left[ (N^2 + D^2) {\cal A}_{\rm u} (n) + 2 ND {\cal A}_{\rm t} 
(n) \right]\, q^{{1 \over 2\alpha'} ( nR)^2 } \,,
\ee
with
\ba
{\cal A}_{\rm u} (0) &=& {v_2 \over \tau_2 \, \eta^4} \sum_{\alpha ,\beta}
{\textstyle{1\over 2}} \, \eta_{\alpha \beta} \, {\vartheta^4 \left[
{\alpha \atop \beta} \right] \over \eta^4} \,,
\nonumber \\
{\cal A}_{\rm u} (n) &=& \sum_{\alpha , \beta} {\textstyle{1\over 2}}\,
\eta_{\alpha \beta} {\vartheta^2 \left[ {\alpha \atop \beta} \right]
\over \eta^2} \, \prod_{i=1,2}  e^{- 2 \pi i \gamma_i n (2 \beta-1)} 
{ \vartheta \left[ {\alpha + 2\gamma_i n
\atop \beta} \right] \over i \ \vartheta \left[ {1/2 + 2 \gamma_i n \atop 1/2}
\right]} \,,
\nonumber \\
{\cal A}_{\rm t} (0) &=& \sum_{\alpha ,\beta}
{\textstyle{1\over 2}} \, \eta_{\alpha \beta} e^{2i\pi\alpha}
\, {\vartheta^2 \left[
{\alpha \atop \beta} \right] \over \eta^2} \, {\vartheta^2 \left[
{\alpha + 1/2 \atop \beta} \right] \over \vartheta^2 \left[ {0 \atop 1/2}
\right]} \,,
\nonumber \\
{\cal A}_{\rm t} (n) &=& \sum_{\alpha , \beta} {\textstyle{1\over 2}}\,
\eta_{\alpha \beta} e^{2i\pi\alpha}
{\vartheta^2 \left[ {\alpha \atop \beta} \right]
\over \eta^2} \, \prod_{i=1,2}  e^{- 2 \pi i \gamma_i n (2 \beta-1)} 
{\vartheta \left[ {\alpha +1/2+ 2\gamma_i n
\atop \beta} \right] \over \vartheta \left[ {2 \gamma_i n \atop 1/2}
\right]} \,,
\ea
and
\be
{\cal M} = -{\textstyle{1\over 2}} \, {v_5 \over \tau_2^{5/2}\, \eta^4}
\sum_n (N+D) \, {\cal M} (n) \, q^{{1\over 2 \alpha '} (nR)^2} \,,
\ee
with
\ba
{\cal M} (0) &=& {v_2 \over \tau_2 \, \eta^4} \sum_{\alpha , \beta}
{\textstyle{1\over 2}} \, \eta_{\alpha \beta}\, {\vartheta^4 \left[
{\alpha \atop \beta}\right] \over \eta^4}\,,
\nonumber \\
{\cal M} (n) &=& \sum_{\alpha , \beta} 
{\textstyle{1\over 2}} \, \eta_{\alpha \beta}\, {\vartheta^2 \left[
{\alpha \atop \beta}\right] \over \eta^2} \prod_{i=1,2}
 e^{- 2 \pi i \gamma_i n (2 \beta-1)}
{ \vartheta \left[ { \alpha + 2\gamma_i n \atop \beta - \gamma_i n} \right]
\over i \ \vartheta \left[ {1/2 + 2\gamma_i n \atop 1/2 - \gamma_i n} \right]}
 \,.
\ea
The Chan-Paton gauge group is ${\rm SO} (N) \times {\rm SO} (D)$, where
the effective number of branes is, as usual, fixed demanding that the
final configuration be neutral and results in $N=D=8$. 
Actually, for the non-standard dependence
on the tree-level proper-time $\ell$ of the zero-modes contributions 
to $\tilde{\cal K}$, $\tilde{\cal A}$ and $\tilde{\cal M}$,
the explicit calculation of NS-NS and R-R tadpoles for this model 
present the same difficulties (and solutions) of section 5. One can
also verify that, in the limit $\gamma_i \to {1\over 2}$, both 
${\cal A}$ and ${\cal M}$ consistently reduce to those of \cite{ssopen}.

\vskip 24pt

\noindent
{\bf Acknowledgements.} We are grateful to Augusto Sagnotti for useful
discussions. A.C. and E.D. thank the Physics Department of the 
University of Rome ``Tor Vergata'' for the warm hospitality
extended to them. Work supported in part
by the RTN European Program HPRN-CT-2000-00148. 

\appendix

\section{Boundary states for Melvin orientifolds}

In this appendix we introduce the boundary states \cite{boundary} for 
our Melvin orientifolds.

\bigskip\noindent
{\it Melvin model.}
The crosscap state $|C,\eta\rangle$ is defined by
\ba
\left(X^\mu(0,\sigma+\pi)-X^\mu(0,\sigma)\right) \; |C,\eta\rangle &=& 0 \, , 
\nonumber \\
\left(\partial_\tau X^\mu(0,\sigma+\pi)+\partial_\tau
X^\mu(0,\sigma)\right) \; |C,\eta\rangle &=&0 \, , 
\nonumber\\
\left(\psi_+(0,\sigma)+i\eta \psi_-(0,\sigma+\pi)\right) \; |C,\eta
\rangle &=& 0 \, ,
\label{o10}
\ea
with $\eta=\pm 1$.
In terms of the Laurent modes, these equations translate into 
\ba
\pi n R \quad \left( {\rm mod}\ 2 \pi R\right)\; |C,\eta\rangle &=& 0 \,, 
\nonumber \\
\left({k \over R} - {\gamma \over R} (J+\tilde J)\right) \; |C,\eta
\rangle &=& 0 \, , 
\nonumber \\ 
\left(y_m - (-1)^m \tilde y_m^\dagger \right) \; |C,\eta\rangle &=& 0 \, , 
\nonumber \\
\left(y^\dagger_m- (-1)^m \tilde y_m \right) \; |C,\eta\rangle &=&0 \, , 
\nonumber \\
\left(\psi_m +i (-1)^m e^{i\epsilon\pi}
\eta \tilde \psi_{m}^\dagger \right) \; |C,\eta\rangle &=& 0 \, ,
\nonumber \\
\left( \psi_m^\dagger  +i (-1)^m e^{-i\epsilon\pi}\eta \ 
\tilde \psi_{m} \right) \; |C,\eta\rangle &=& 0 \,,
\nonumber \\
\left(\psi_0+i\eta \tilde \psi_0 \right) |C,\eta\rangle &=& 0\,,
\label{o11}
\ea
for the compact $y$ coordinate\footnote{The mode expansion of $\psi_+$
in the NS sector reads $\sum_{m=1}^{\infty}\psi_m e^{-i(m-1/2)\sigma_+}+ {\rm
h. c.}$ and in the R sector $\sum_{m=1}^{\infty}\psi_m e^{-im\sigma_+}
+{\rm h.c.} +\psi_0$.}, and 
\ba
\left( a_m - (-1)^m \tilde a_m^{\dagger} \right) \; |\psi_n,\eta\rangle &=&0 
\,, 
\nonumber \\
\left(\tilde b_m- (-1)^m b_m^{\dagger}\right) \; |\psi_n,\eta\rangle
&=&0 \, , 
\nonumber \\
\left( a_m^\dagger - (-1)^m \tilde a_m \right) \; |\psi_n,\eta\rangle
&=&0 \, , 
\nonumber \\
\left(\tilde b_m^\dagger - (-1)^m b_m \right) \; |\psi_n,\eta\rangle
&=&0 \, , 
\nonumber\\
\left(\mu_m^\dagger+ i (-1)^m \eta \ \ e^{-i \epsilon \pi} \tilde \mu_{m}
\right) \; |\psi_n,\eta\rangle &=& 0 \,, 
\nonumber \\
\left(\tilde \zeta_m^\dagger- i (-1)^m \eta  e^{-i \epsilon \pi} \  
\zeta_{m}\right) \; |\psi_n,\eta\rangle &=& 0 \, , 
\nonumber \\
\left( \mu_m + i (-1)^m \eta \ \ e^{i \epsilon \pi} 
{\tilde \mu}_{m}^{\dagger} \right) 
\; |\psi_n,\eta\rangle &=& 0 \,, 
\nonumber \\ 
\left(\tilde \zeta_m - i (-1)^m \eta  e^{i \epsilon \pi} \  
\zeta_{m}^{\dagger} \right) 
\; |\psi_n,\eta\rangle &=&0 \,,
\label{o14}
\ea
for the non-compact (twisted) $Z_0$ coordinate, with $\epsilon=0$ in
the R sector and ${1\over 2}$ in the NS one.
In eq. (\ref{o11}), the last line applies  to R sector only. Moreover,
the first line implies that only {\it even} windings couple
to the boundary state $|C,\eta\rangle$. Since $\gamma $ is irrational, 
from the second condition we get that both the KK
momentum $k$ and the total angular momentum $J+\tilde J$ must vanish.
Actually, $ (J+\tilde J) \; |C, \eta\rangle = 0$
together with (\ref{angm}), implies that one set of closed-string 
Landau levels couples to the boundary state. This indeed matches with the 
tree-level amplitude (\ref{o7}), that displays manifestly the tree-level
propagation of Landau levels between the two O8 planes.

The solution for the crosscap state is then
\be
|C,\eta\rangle =\sum_{n\ {\rm even}} {\cal N}_n\ |n\rangle\otimes 
\exp\left\{ \sum_{m=1}^{\infty}
(-1)^m ({y_m}^\dagger{\tilde y_m}^\dagger-i
\eta  e^{i \epsilon \pi} \ \psi_{m}^\dagger\tilde
\psi_{m}^\dagger) \right\}|0\rangle \otimes|\psi_n,\eta\rangle \, , 
\label{o13}
\ee
with
\be
|\psi_n,\eta\rangle = \exp \left\{ 
\sum_{m=0}^\infty (-1)^m (\tilde b_m^\dagger b_m^\dagger
+i \eta e^{i \epsilon \pi} \zeta_m^\dagger\tilde \zeta_m^\dagger) 
+ \sum_{m=1}^\infty
(-1)^m (a_m^\dagger \tilde a_m^\dagger - 
i \eta  e^{-i  \epsilon \pi} \mu^\dagger_{m} \tilde \mu^\dagger_{m}) 
\right\} \; |0\rangle
\, , \label{o15}
\ee
and 
\be
{\cal N}_{n}=2\pi T'_9 {1+(-1)^n \over 2} \sqrt{\alpha' v_{8}
\cot (\pi \gamma n/2)}  \label{o04}
\ee
a normalization constant, determined from eq. ({\ref{o7}), where $T'_9 =
32 T_9$ id the O9 tension.

A nice check of the conditions (\ref{o14}) is their invariance under 
the orientifold $\Omega$ involution. In the tree-level channel, 
$\Omega$ acts as
\be
\Omega \; y (\tau) \; \Omega^{-1} = y (-\tau) \, , \qquad 
\Omega \; Z_0^{(\nu)} (\tau) \; \Omega^{-1} = Z_0^{(\nu)} (-\tau) \, . 
\label{o16}
\ee
In terms of the oscillator modes, (\ref{o16}) translate into
\be
\Omega \; y_m \; \Omega^{-1} = {\tilde y}_m^{\dagger} \,, \quad
\Omega \; a_m \; \Omega^{-1} = {\tilde a}_m^{\dagger} \,, \quad 
\Omega \; {\tilde b}_m \; \Omega^{-1} = b_m^{\dagger} \,, \label{o17}
\ee
and similar ones for the RNS fermions. 
Then, the $\Omega$-invariance of eqs. (\ref{o11}) and (\ref{o14}) comes  
naturally from (\ref{o16}) and from $\Omega |C, \eta\rangle =|C, 
\eta\rangle $.  Notice in particular that eqs.
(\ref{o17}) imply $ \Omega \; J \; \Omega^{-1} = - {\tilde J}$, which
selects one set of Landau levels, thus allowed to couple to 
the O-planes.  

In the boundary-state formalism, the tree-level Klein-bottle amplitude 
(\ref{o7}) is then given by
\be
\langle C | e^{-2 \pi \ell H} |C \rangle = 
\sum_n \langle C | e^{-2 \pi \ell [N - 2
\nu J + (n^2 R^2/2 \alpha') + a]} |C \rangle \,. \label{o18}
\ee   
where $|C\rangle = |C,+\rangle \pm |C,-\rangle$ in the R-R (NS-NS) sector
is the GSO projected crosscap state.

\bigskip\noindent
{\it Dual Melvin orientifold.}
In the dual Melvin model the crosscap state $|C,\eta\rangle$ is defined by
\ba
\left( y(0,\sigma+\pi)+y(0,\sigma) \right) \; |C,\eta\rangle &=& 0 \, , 
\nonumber \\  
\left( \partial_\tau y(0,\sigma+\pi)-\partial_\tau y(0,\sigma) \right) 
\; |C,\eta\rangle &=& 0 \,, 
\nonumber \\
\left(\psi_+(0,\sigma) - i \eta \psi_-(0,\sigma+\pi) \right)\; |C,\eta\rangle
&=& 0 \, ,
\nonumber \\
\left( Z^{\dagger} (0,\sigma+\pi)-Z(0,\sigma) \right) \; |C,\eta\rangle 
&=& 0 \, , 
\nonumber \\
\left( \partial_\tau Z^{\dagger}(0,\sigma+\pi)+\partial_\tau
Z(0,\sigma) \right) \; |C,\eta\rangle &=& 0 \, , 
\nonumber\\
\left( \lambda_{+} (0,\sigma)+ i \eta \lambda_{-}^{\dagger} (0,\sigma+\pi) 
\right) \; |C,\eta\rangle &=& 0 \,.
 \label{d18}
\ea
It is actually more convenient to write
the crosscap state for the twisted (but free) $Z_0$ coordinate,
though, in this case, some care is needed in implementing the
identification (\ref{one}).
A generic state $|C\rangle$ is in fact a linear combination of states
of the form $ e^{ip_y y} |J,\bar J,\dots\rangle $, 
where $|J,\bar J,\dots\rangle$ is an 
eigenstate of the total angular momentum $J+\bar J$.
The invariance under (\ref{one}), thus implies
that $ p_y R+ \gamma (J+\bar J)=k $, with $k$ an integer.
As a result the crosscap state can be put in the form
\be
|C,\eta\rangle =\sum_k {\cal N}_k 
e^{-iy_0(J+\bar J)\gamma/R} |k\rangle \otimes|\psi_k,\eta\rangle \, ,
\label{d181}
\ee
where $y_0$ is the generator of translations in momentum space,
and $|\psi_k ,\eta \rangle$ encodes the contributions of the remaining 
modes. Furthermore, the zero-mode part of the first equation in 
(\ref{d18}) 
\be
(2y_0-2p \pi) \; |C,\eta\rangle =0 \qquad (p\in\bb{Z})\,, \label{d180}
\ee
implies that $|C\rangle$ has no windings and fixes
${\cal N}_k ={\cal N}_p \, e^{-i k p\pi}$. 

Using (\ref{d181}) and the relation $ e^{iy_0(J+\bar J)\gamma/R} 
\; Z_0(\sigma) \; e^{-iy_0(J+\bar J)\gamma/R}=
e^{iy_0\gamma/R}Z_0(\sigma) $, one can then recast the fourth and fifth
equations in (\ref{d18})
\ba
\left( Z_0(\sigma)e^{-i\gamma y_0/R}-Z_0^\dagger(\sigma+\pi)
e^{i\gamma y_0/R}\right) \; |C,\eta\rangle &=& 0 \,, 
\nonumber\\
\left(\partial_\tau Z_0(\sigma)e^{-i\gamma y_0/R}+\partial_\tau
Z_0^\dagger(\sigma+\pi)
e^{i\gamma y_0/R} \right) \; |C,\eta\rangle &=&0 \, 
\ea
in the form
\ba
\left( Z_0(\sigma)- Z_0(\sigma+\pi)^\dagger \right) \; |\psi_k ,\eta
\rangle &=&0 
\, , 
\nonumber\\
\left( \partial_\tau Z_0(\sigma)+ 
\partial_\tau Z_0(\sigma+\pi)^\dagger \right) \; |\psi_k ,\eta\rangle &=& 0 \,,
\label{d120}
\ea
whose zero-mode part implies that 
$|\psi_k ,\eta \rangle $ has zero eigenvalues for $Y_0$ and $P_{X_0}$.
For the oscillators one finds the conditions
\ba
\left( y_m + (-1)^m \tilde y_m^\dagger \right) \; |\psi_k,\eta\rangle 
&=& 0 \,, 
\nonumber \\ 
\left(y^\dagger_m+ (-1)^m \tilde y_m\right) \; |\psi_k,\eta\rangle &=&0 \, , 
\nonumber \\
\left(\psi_m - i (-1)^m \eta  e^{i \epsilon \pi} 
\tilde \psi_{m}^\dagger\right) \; |\psi_k,\eta\rangle  &=& 0 \,, 
\nonumber \\
\left(\psi_m^\dagger - i (-1)^m \eta  
e^{-i \epsilon \pi} \tilde \psi_{m}\right) \; |\psi_k,\eta\rangle &=& 0 \,, 
\nonumber \\
\left( a_m - (-1)^m  \tilde b_m^{\dagger} \right) \; |\psi_k,\eta\rangle &=& 0 
\,, 
\nonumber \\
\left( \tilde a_m- (-1)^m   b_m^{\dagger}\right) \;
|\psi_k,\eta\rangle &=& 0 \, , 
\nonumber \\
\left( a_m^\dagger - (-1)^m   \tilde b_m \right) \; 
|\psi_k,\eta\rangle &=&0 \,,
\nonumber \\
\left( \tilde a_m^\dagger - (-1)^m   b_m \right) \;
|\psi_k,\eta\rangle &=& 0 \, , 
\nonumber\\
\left( \mu_m+ i (-1)^m \eta  e^{i \epsilon \pi} \ 
{\tilde \zeta}_{m}^{\dagger} \right) \; |\psi_k,\eta\rangle &=& 0 \, ,
\nonumber \\  
\left( \tilde \mu_m^\dagger- i(-1)^m \eta  e^{i \epsilon \pi} 
\ \zeta_{m} \right) \; |\psi_k,\eta\rangle &=& 0 \, , 
\nonumber \\
\left(\zeta_0^\dagger +i \eta  
\tilde \zeta_0^\dagger\right) |\psi_k,\eta\rangle &=& 0 \, ,
\nonumber \\ 
\left(\zeta_0 +i \eta   \tilde \zeta_0 \right)
|\psi_k,\eta\rangle &=& 0 \, .  \label{d19}
\ea

The solution to eqs. (\ref{d120})--(\ref{d19}) then reads
\ba
|C,\eta\rangle &=&\sum_{p=0,1} \sum_{k} {\cal N}_p e^{-ikp\pi}
 \e^{-iy_0\gamma(J+\bar J)/R} \; |k\rangle \otimes 
\int dP_{Y_0} \ |P_{X_0}=0,P_{Y_0}\rangle
\nonumber\\
& &\otimes
\exp \left\{- \sum_{m=1}^{\infty}
(-1)^m ({y_m}^\dagger{\tilde y_m}^\dagger + i \eta  
e^{i \epsilon \pi} \psi_{m}^\dagger\tilde
\psi_{m}^\dagger) \right\}\; |0\rangle
\nonumber\\
& &\otimes \exp\left\{\sum_{m=1}^\infty (-1)^m 
\left[  \tilde b_m^\dagger a_m^\dagger
+  b_m^\dagger \tilde a_m^\dagger -i
\eta  e^{i \epsilon \pi}   \mu_{m}^{\dagger} \tilde
\zeta_{m}^{\dagger}+ i \eta   e^{-i \epsilon \pi} 
{\tilde \mu}_{m}^{\dagger} \zeta_{m}^{\dagger} \right] \right\} \; 
|0\rangle \, , \label{o130}
\ea
where $k$ is the KK momentum and ${\cal N}_p$ is a normalization
constant fixed by the transverse-channel Klein-bottle
amplitude. In fact, for the bosonic fields and after a Poisson 
resummation in $k$,
\ba
\langle C|e^{-\pi \ell H}|C\rangle &=& \sum_{p,p',\tilde k} {\cal N}_{p'}^*
{\cal N}_p
\sqrt{{2R^2 \over \ell\alpha'}} e^{-{\pi R^2 \over 2\ell\alpha'} (2\tilde k
-(p-p'))^2}
\nonumber\\
&&\times \langle Y_0=0,P_{X_0}=0| e^{i(2\tilde k-(p-p'))\gamma\pi L}
e^{-{1\over 2} \pi
\ell\alpha'(P_{X_0}^2+P_{Y_0}^2)} |P_{X_0}=0,Y_0=0\rangle 
\nonumber\\
&& \times \prod_{m=1}^{\infty} { 1\over(1-e^{-2\pi \ell m} 
e^{i(2\tilde k-(p-p'))\gamma\pi})(1-e^{-2\pi \ell m} 
e^{-i(2\tilde k-(p-p'))\gamma\pi})} \,,
\ea
where the first and second lines receive contributions from 
the zero modes while the third from the $Z_0$ oscillators.
The matrix element in the second line gives
\be
{1 \over |\sin{(2\tilde k-(p-p'))\gamma\pi}|} \qquad {\rm if} \ 
2 \tilde k-(p-p')\neq 0 \ ,
\ee
and $(L/2 \pi) \sqrt{(2/\ell\alpha')}$ otherwise, where $L$ is the length 
in the $Y_0$ direction. Therefore, comparison with the transverse
amplitude (\ref{pippo}) fixes ${\cal N}_{p=0}={\cal N}_{p=1}= {\cal N}=
(1/\sqrt{R \alpha'^3})$. 
Notice that, in the direct-channel, even (odd) winding modes are
associated to the terms with $p=p'$ ($p\not= p'$).

A geometrical picture of the O-planes is nicely spelled out from
the zero-mode contributions to the crosscap state in the position
representation
\be
c(y, X_0,Y_0)= {\cal N} \sum_{p=0,1}\sum_k
e^{ik(y-p\pi R)/R} \ 
\delta \left( \sin \left({\gamma y\over R}\right) X_0 -
\cos \left({\gamma y\over R}\right)Y_0 \right) \,.
\ee
We therefore have an infinity of orientifold planes localized at
$(y=p \pi R, \phi_0= p \gamma\pi)$. 

Finally, boundary states can be build following similar routes. 
Their bosonic part is given by 
\ba
|B\rangle &=& {1 \over 8} \sum_k {\cal N} e^{-ikr_0/R} \, 
e^{-iy_0\gamma(J+\bar J)/R}
\; |k\rangle \otimes \exp\left\{-\sum_{m=1}^{\infty}
{y_m}^\dagger{\tilde y_m}^\dagger \right\} \; |0\rangle
\nonumber\\
&& \otimes |b\rangle \otimes \exp \left\{ -\sum_{m=1}^{\infty} 
(a_m^\dagger\tilde b_m^\dagger
+b_m^\dagger\tilde a_m^\dagger) \right\} |0\rangle \,, \label{bo2}
\ea
with $r_0$ the position of the brane, and 
$|b\rangle$ is the zero modes contribution in the $Z_0$ plane which
reads
\be
|b\rangle =\int dP_{Y_0} \ |P_{X_0}=0,P_{Y_0}\rangle \, , \label{bo3} 
\ee
encoding the zero-mode contribution in the $Z_0$ plane
(for simplicity we have assumed $\phi_0=0$).
The transverse-channel annulus and M{\"o}bius amplitudes are then given by
$\int d\ell \langle B|e^{-\pi \ell H}|B\rangle$
and $2 \; \int d\ell \langle C|e^{-\pi \ell H}|B\rangle$.


\begin{thebibliography}{99}


\bibitem{Melvin} M.A. Melvin, 
``Pure magnetic and electric geons,''
Phys. Lett. 8 (1964) 65.

\bibitem{Gibbons:1987wg}
G.~W.~Gibbons and D.~L.~Wiltshire,
``Space-time as a membrane in higher dimensions,''
Nucl.\ Phys.\ B287 (1987) 717
[arXiv:hep-th/0109093];
G.~W.~Gibbons and K.~Maeda,
``Black holes and membranes in higher dimensional theories with dilaton
fields,''
Nucl.\ Phys.\ B298 (1988) 741.

\bibitem{dowker} 
F.~Dowker, J.~P.~Gauntlett, D.~A.~Kastor and J.~Traschen,
``Pair creation of dilaton black holes,''
Phys.\ Rev.\ D49 (1994) 2909
[arXiv:hep-th/9309075],
``On pair creation of extremal black holes and Kaluza-Klein monopoles,''
Phys.\ Rev.\ D50 (1994) 2662
[arXiv:hep-th/9312172],
``The decay of magnetic fields in Kaluza-Klein theory,''
Phys.\ Rev.\ D52 (1995) 6929
[arXiv:hep-th/9507143],
``Nucleation of $P$-branes and fundamental strings,''
Phys.\ Rev.\ D53 (1996) 7115
[arXiv:hep-th/9512154].

\bibitem{dm} E.~Dudas and J.~Mourad,
``D-branes in string theory Melvin backgrounds,''
Nucl.\ Phys.\ B622 (2002) 46
[arXiv:hep-th/0110186].

\bibitem{ss}
J.~Scherk and J.~H.~Schwarz,
``How to get masses from extra dimensions,''
Nucl.\ Phys.\ B153 (1979) 61.

\bibitem{ssclosed}
R.~Rohm,
``Spontaneous supersymmetry breaking in supersymmetric string theories,''
Nucl.\ Phys.\ B237 (1984) 553;
S.~Ferrara, C.~Kounnas and M.~Porrati,
``General dimensional reduction of ten-dimensional supergravity and
superstring,''
Phys.\ Lett.\ B181 (1986) 263,
``Superstring solutions with spontaneously broken four-dimensional
supersymmetry,''
Nucl.\ Phys.\ B304 (1988) 500,
``$N=1$ superstrings with spontaneously broken symmetries,''
Phys.\ Lett.\ B206 (1988) 25;
C.~Kounnas and M.~Porrati,
``Spontaneous supersymmetry breaking in string theory,''
Nucl.\ Phys.\ B310 (1988) 355;
S.~Ferrara, C.~Kounnas, M.~Porrati and F.~Zwirner,
``Superstrings with spontaneously broken supersymmetry and their effective
theories,''
Nucl.\ Phys.\ B318 (1989) 75;
C.~Kounnas and B.~Rostand,
``Coordinate dependent compactifications and discrete symmetries,''
Nucl.\ Phys.\ B341 (1990) 641;
I.~Antoniadis and C.~Kounnas,
``Superstring phase transition at high temperature,''
Phys.\ Lett.\ B261 (1991) 369;
E.~Kiritsis and C.~Kounnas,
``Perturbative and non-perturbative partial supersymmetry breaking:  $N = 4
\to N = 2 \to N = 1$,''
Nucl.\ Phys.\ B503 (1997) 117
[hep-th/9703059];
C.~A.~Scrucca and M.~Serone,
``A novel class of string models with Scherk-Schwarz 
JHEP 0110 (2001) 017
[arXiv:hep-th/0107159].

\bibitem{Russo:1995tj}
J.~G.~Russo and A.~A.~Tseytlin,
``Exactly solvable string models of curved space-time backgrounds,''
Nucl.\ Phys.\ B449 (1995) 91
[hep-th/9502038],
``Magnetic flux tube models in superstring theory,''
Nucl.\ Phys.\ B461 (1996) 131
[hep-th/9508068].

\bibitem{Costa:2001nw}
M.~S.~Costa and M.~Gutperle,
``The Kaluza-Klein Melvin solution in M-theory,''
JHEP 0103 (2001) 027
[hep-th/0012072].

\bibitem{Gutperle:2001mb}
M.~Gutperle and A.~Strominger,
``Fluxbranes in string theory,''
JHEP 0106 (2001) 035 
[hep-th/0104136].

\bibitem{Costa:2001if}
M.~S.~Costa, C.~A.~Herdeiro and L.~Cornalba,
``Flux-branes and the dielectric effect in string theory,''
hep-th/0105023.

\bibitem{Emparan:2001rp}
R.~Emparan,
``Tubular branes in fluxbranes,''
Nucl.\ Phys.\ B610 (2001) 169
[hep-th/0105062].

\bibitem{Saffin:2001jg}
P.~M.~Saffin,
``Fluxbranes from p-branes,''
hep-th/0105220.

\bibitem{Brecher:2001xj}
D.~Brecher and P.~M.~Saffin,
``A note on the supergravity description of dielectric branes,''
Nucl.\ Phys.\ B613 (2001) 218
[arXiv:hep-th/0106206].

\bibitem{Suyama:2001bn}
T.~Suyama,
``Closed string tachyons in non-supersymmetric heterotic theories,''
JHEP 0108 (2001) 037
[hep-th/0106079],
``Melvin background in heterotic theories,''
hep-th/0107116,
``Properties of string theory on Kaluza-Klein Melvin background,''
hep-th/0110077.

\bibitem{Adams:2001sv}
A.~Adams, J.~Polchinski and E.~Silverstein,
``Don't panic! Closed string tachyons in ALE space-times,''
hep-th/0108075.

\bibitem{Uranga:2001dx}
A.~M.~Uranga,
``Wrapped fluxbranes,''
hep-th/0108196.

\bibitem{Russo:2001na}
J.~G.~Russo and A.~A.~Tseytlin,
``Supersymmetric fluxbrane intersections and closed string tachyons,''
hep-th/0110107.

\bibitem{Takayanagi:2001jj}
T.~Takayanagi and T.~Uesugi,
``Orbifolds as Melvin geometry,'' hep-th/0110099.

\bibitem{tu2}
T.~Takayanagi and T.~Uesugi,
``D-branes in Melvin background,''
JHEP 0111 (2001) 036
[arXiv:hep-th/0110200],
``Flux stabilization of D-branes in NS-NS Melvin background,''
arXiv:hep-th/0112199.

\bibitem{eg}
R.~Emparan and M.~Gutperle,
``From p-branes to fluxbranes and back,''
JHEP 0112 (2001) 023
[arXiv:hep-th/0111177].

\bibitem{my}
Y.~Michishita and P.~Yi,
``D-brane probe and closed string tachyons,''
arXiv:hep-th/0111199.

\bibitem{dghm}
J.~R.~David, M.~Gutperle, M.~Headrick and S.~Minwalla,
``Closed string tachyon condensation on twisted circles,''
arXiv:hep-th/0111212.

\bibitem{cargese} A. Sagnotti, in: Cargese '87, Non-Perturbative Quantum
Field Theory, eds. G. Mack et al. (Pergamon Press, Oxford, 1988) p. 521; 
M.~Bianchi and A.~Sagnotti,
``On the systematics of open string theories,''
Phys.\ Lett.\ B247 (1990) 517,
``Twist symmetry and open string Wilson lines,''
Nucl.\ Phys.\ B361 (1991) 519;
G.~Pradisi and A.~Sagnotti,
``Open string orbifolds,''
Phys.\ Lett.\ B216 (1989) 59;
M.~Bianchi, G.~Pradisi and A.~Sagnotti,
``Toroidal compactification and symmetry breaking in open string theories,''
Nucl.\ Phys.\ B376 (1992) 365.

\bibitem{ad} E.~Dudas,
``Theory and phenomenology of type I strings and M-theory,''
Class.\ Quant.\ Grav.\ 17 (2000) R41
[arXiv:hep-ph/0006190].

\bibitem{review} C.~Angelantonj and A.~Sagnotti,
``Open strings,''
arXiv:hep-th/0204089.

\bibitem{ads1}
I.~Antoniadis, E.~Dudas and A.~Sagnotti,
``Supersymmetry breaking, open strings and M-theory,''
Nucl.\ Phys.\ B544 (1999) 469
[hep-th/9807011].

\bibitem{ssopen}
I.~Antoniadis, G.~D'Appollonio, E.~Dudas and A.~Sagnotti,
``Partial breaking of supersymmetry, open strings and M-theory,''
Nucl.\ Phys.\ B553 (1999) 133
[hep-th/9812118].

\bibitem{mag} C.~Angelantonj, I.~Antoniadis, E.~Dudas and A.~Sagnotti,
``Type-I strings on magnetised orbifolds and brane transmutation,''
Phys.\ Lett.\ B489 (2000) 223
[arXiv:hep-th/0007090];
C.~Angelantonj and A.~Sagnotti,
``Type-I vacua and brane transmutation,''
arXiv:hep-th/0010279.

\bibitem{bcw}
C.~Bachas, N.~Couchoud and P.~Windey,
``Orientifolds of the 3-sphere,''
JHEP 0112 (2001) 003
[arXiv:hep-th/0111002].

\bibitem{augusto}
A.~Sagnotti,
``Some properties of open string theories,''
arXiv:hep-th/9509080 .

\bibitem{dm3}
E.~Dudas and J.~Mourad,
``D-branes in non-tachyonic 0B orientifolds,''
Nucl.\ Phys.\ B598 (2001) 189
[arXiv:hep-th/0010179].

\bibitem{witten} E.~Witten,
``Instability Of The Kaluza-Klein Vacuum,''
Nucl.\ Phys.\ B195 (1982) 481.

\bibitem{boundary}
C.~G.~Callan, C.~Lovelace, C.~R.~Nappi and S.~A.~Yost,
``Adding holes and crosscaps to the superstring,''
Nucl.\ Phys.\ B293 (1987) 83;
J.~Polchinski and Y.~Cai,
``Consistency of open superstring theories,''
Nucl.\ Phys.\ B296 (1988) 91;
N.~Ishibashi,
``The boundary and crosscap states in conformal field theories,''
Mod.\ Phys.\ Lett.\ A4 (1989) 251.


\end{thebibliography}
\end{document}